\documentclass[final,5p,times,twocolumn,authoryear]{elsarticle}

\usepackage{graphicx}
\usepackage{txfonts}
\usepackage{lscape}             
\usepackage{placeins}           

\usepackage{amsmath,mathrsfs,graphicx,float,latexsym,url}
\usepackage{epstopdf,ragged2e}
\usepackage{natbib,url}
\bibpunct{(}{)}{;}{a}{}{,}
\usepackage{hyperref}

\usepackage[usenames,dvipsnames]{color}
\definecolor{darkred}{rgb}{0.5,0,0}
\definecolor{darkgreen}{rgb}{0,0.5,0}
\definecolor{darkblue}{rgb}{0,0,0.5}
\definecolor{prussian}{rgb}{0.0, 0.19, 0.33}
\definecolor{richelectricblue}{rgb}{0.03, 0.57, 0.82}
\definecolor{teal}{rgb}{0.0, 0.5, 0.5}
\definecolor{mediumseagreen}{rgb}{0.24, 0.7, 0.44}
\definecolor{lust}{rgb}{0.9, 0.13, 0.13}
\definecolor{ballblue}{rgb}{0.13, 0.67, 0.8}
\definecolor{darkcyan}{rgb}{0.0, 0.55, 0.55}
\definecolor{mountainmeadow}{rgb}{0.19, 0.73, 0.56}
\definecolor{palecarmine}{rgb}{0.69, 0.25, 0.21}
\definecolor{richcarmine}{rgb}{0.84, 0.0, 0.25}
\definecolor{tangelo}{rgb}{0.98, 0.3, 0.0}
\definecolor{venetian}{rgb}{0.784,0.031,0.082}
\definecolor{bdfrance}{rgb}{0.192,0.549,0.906}
\usepackage{booktabs}
\usepackage{mathtools}
\usepackage{amsbsy}
\usepackage{bm}
\usepackage{float}
\usepackage{array}
\usepackage{tabularx}
\usepackage{multirow}
\usepackage{verbatim}

\newcommand{\Bs}{B_{\star}}
\newcommand{\RA}{R_{\rm A}}
\newcommand{\rnu}{r_{\nu}}
\newcommand{\kms}{km\,s$^{-1}$}

\usepackage{anyfontsize} 
\DeclareFontShape{OML}{cmm}{b}{it}{
   <5> <6> <7> <8> <9> <10> <10.95> <12> <14.4> <17.28> <20.74> <24.88>
   cmmib10
}{}

\definecolor{darkgreen}{RGB}{0, 180, 120}

\usepackage{amssymb}
\usepackage{lipsum}
\usepackage{macros} 
\usepackage{amsmath}
\usepackage{booktabs}
\usepackage{array}
\usepackage{tabularx}
\usepackage{multirow}
\usepackage{verbatim}
\usepackage[percent]{overpic}
\usepackage[dvipsnames]{xcolor}

\journal{High Energy Astrophysics}

\begin{document}

\begin{frontmatter}

\title{Transient discs around isolated accreting neutron stars}

\author[first,second]{Marina D. Afonina\texorpdfstring{\corref{cor1}}{}}
\cortext[cor1]{Corresponding author}
\ead{afonina.md19@physics.msu.ru}
\author[third,first]{Anton V. Biryukov}
\author[first]{Sergei B. Popov}
\author[fourth]{Arthur G. Suvorov}
\affiliation[first]{organization={Sternberg Astronomical Institute,  Lomonosov Moscow State University},
            addressline={13 Universitetskij pr.}, 
            city={Moscow},
            postcode={119234}, 
            country={Russia}}
\affiliation[second]{organization={Department of Physics, Lomonosov Moscow State University},
            addressline={1/2 Leninskie Gory}, 
            city={Moscow},
            postcode={119991},
            country={Russia}}
\affiliation[third]{organization={The Raymond and Beverly Sackler School of Physics and Astronomy, Tel Aviv University}, 
            addressline={55 Chaim Levanon St.}, 
            city={Tel Aviv},
            postcode={6997801},
            country={Israel}}
\affiliation[fourth]{organization={Theoretical Astrophysics, IAAT, University of Tübingen}, 
            addressline={Auf der Morgenstelle 10}, 
            city={Tübingen},
            postcode={D-72076},
            country={Germany}}
\begin{abstract}
 
\noindent Mature, isolated neutron stars can accrete from the interstellar medium. 
 Due to turbulence, the accreted gas can have a substantial angular momentum and form a disc around the compact object. 
 In this paper, we perform a population synthesis of isolated neutron stars in the Milky Way that specifically tracks the possibility of disc formation, which typically requires a low spatial velocity of the compact object ($\lesssim 40$~km~s$^{-1}$). 
 In general, weak magnetic fields favour disc formation as the magnetosphere occupies a smaller volume. 
 Still, even in the case of fields decaying exponentially over a characteristic timescale of $\sim 1.5$~Gyr, we find that for several realistic models of propeller spin-down, only a small fraction of accretors (down to $\sim 0.02$\%) attain discs. 
 However, disc-accreting isolated neutron stars are relatively numerous among the brightest sources.
 We estimate that their number can reach a few hundred at X-ray fluxes of $\gtrsim 10^{-14}$~erg~s$^{-1}$~cm$^{-2}$. 
 We speculate that disc-accreting isolated neutron stars can manifest as long-period radio transient sources via electron cyclotron maser emission, predicting spectra and critical conditions for quenching.


\end{abstract}




\end{frontmatter}



\section{Introduction}

 The number of neutron stars (NSs) in the Galaxy is thought to be of order a few hundred million \cite[e.g.,][]{2000PASP..112..297T}. 
 The majority of them are isolated as, even if the progenitor has been a member of a binary (or multiple) system, natal kicks tend to cause them to become gravitationally unbound \citep{2019A&A...624A..66R}. 

 Isolated NSs are observed mainly in their youth as radio pulsars, magnetars, rotating radio transients, and/or central compact objects in supernova (SN) remnants \citep{2023Univ....9..273P, 2023hxga.book..146B, 2024mbhe.confE..55R}. 
 Mature, isolated NSs are more elusive as their luminosity is expected to be quite low due to the gradual exhaustion of thermal, magnetic, and rotational energy. 
 However, the interstellar medium (ISM) provides a reservoir of material available for accretion that could, in principle, fuel these stars even if very old \citep{1970ApL.....6..179O, 1971SvA....14..662S} and an amount of theoretical effort has gone towards understanding such interactions may take place.

 Under optimistic assumptions, accreting isolated NSs (AINSs) can appear as soft X-ray sources with luminosities of order $\sim 10^{31}-10^{32}$~erg~s$^{-1}$ \citep{2026JHEAp..5300643A}. In the early 1990s, just before the launch of the {\it ROSAT} X-ray observatory, it was estimated that the number of detectable AINSs could be about a few thousand \citep{1991A&A...241..107T}.  
However, up to now, there are no even weak X-ray candidates for AINSs \cite[see, e.g.,][]{2011AJ....141..176A, 2024A&A...687A.251K}. 
Several explanations are possible to resolve the discrepancy, which can be roughly separated into two groups: evolutionary aspects and the accretion efficiency. 

 The first (evolutionary) set of proposed solutions suggests that just a small fraction of INSs can reach the stage of accretion. This can be due to large spatial velocities relative to the Galactic potential, which prolong the so-called ejector stage, in which electromagnetic radiation and a strong wind of relativistic particles (produced at the expense of rotational kinetic-energy) prevent external particles from entering the NS magnetosphere. 
 Otherwise, a weak spin-down of the NS at the propeller stage can prevent the onset of accretion on the Hubble time scale. The first option was studied using a population synthesis approach by \cite{2000ApJ...530..896P} and others since, while the second more recently by \cite{2026JHEAp..5300643A}. These studies showed however that, in practice, large kick velocities could reduce the number of AINSs by only a factor of a few, and only very inefficient spin-down during the propeller stage could bring the number of detectable sources below observational limits.

If the efficiency is instead responsible for their lack of observation, AINSs should be numerous but dim. Initially, it was proposed that a low efficiency could again occur due to large kicks \cite[see, e.g.,][for a review]{2000PASP..112..297T}. 
If AINSs belong only to the low-velocity tail of the distribution, however, other reasons might be considered; for example, \cite{2003ApJ...593..472T, 2012MNRAS.420..810T} 
demonstrated via numerical simulations of Bondi accretion that even relatively weak ($\sim 10^8$~G) magnetic fields invite outward-propagating shocks that prevent significant accretion onto the surface. 
In addition, at low rates, in the so-called settling accretion regime \citep{2012MNRAS.420..216S}, captured matter may cool too slowly such that 
the hot material fails to enter the magnetosphere at the Bondi rate \citep{2015MNRAS.447.2817P}. 

However, in some rare cases the accretion efficiency can be increased. One way to boost it is for a disc to form around the compact object. Disc formation, being rare, might not influence the general population of AINSs. Still, disc-accreting INSs can be less elusive and demonstrate interesting properties.
{ Disc formation around an AINS} can happen due to interstellar turbulence, as in that case the material carries a non-negligible angular momentum \citep{2002A&A...381.1000P}. 
To understand how common such a scenario is, and its impact with respect to detectability and the Galactic population of AINSs more generally, we present the results of a population synthesis model in this paper, building on the work of \cite{2026JHEAp..5300643A}.

Aside from X-ray channels, \cite{ferrario25} recently proposed a model for the so-called long-period transients (LPTs) based on AINSs. We advance this proposal, suggesting that INSs with disc accretion can manifest themselves as a sub-class of LPTs and develop theoretical criteria for when an AINS can manifest as such a source.
Importantly, long epochs of interaction with the interstellar medium (ISM) can naturally allow for neutron stars to attain long periods.
This is difficult to achieve with innate mechanisms, though some proposals based on quake-fuelled spindown have been considered in the literature \cite[see][]{2026ApJ..1000...55S}.
The population of AINSs with discs we find -- such that certain resonance and quenching conditions are satisfied -- is consistent with the observed number of LPTs that show no evidence for binarity \citep{2026JHEAp..5200566R}, further motivating their study. 

This paper is organised as follows.
 In the next section, we describe the basics of accretion onto INSs from the interstellar medium and derive some properties of AINSs. Then in Section~\ref{sec:pops} we summarise the model of the population synthesis of Galactic INSs developed by \cite{2026JHEAp..5300643A}. 
 In Section~\ref{sec:results}, we present results of our simulations. 
 Possible appearance of disc-accreting INSs (dAINSs) is discussed in Section~\ref{sec:lprt}. 
 In the final section, we discuss the models we used and the results and present our conclusions.
 
\section{Accretion onto an isolated neutron star from a turbulised interstellar medium}
\label{sec:acc}

Typically, accretion onto isolated NSs from the ISM is described in terms of a quasi-spherical flow with the accretion rate estimated with the \cite{1952MNRAS.112..195B} formula:
 \begin{equation}
     \dot M = \xi \pi R_\mathrm{G}^2 \rho v.
 \end{equation}
 Here, $R_\mathrm{G}=2GM/v^2$ is the gravitational capture (Bondi) radius, where $G$ and $M$ denote Newton's constant and the NS mass, respectively. The parameter $v$ is a characteristic velocity of the external material relative to the NS. In general, $v$ can include the spatial velocity of the NS relative to the ISM $v_\text{rel}$, sound speed $c_\text{s}$, and/or other types of motion (e.g., $ v^2=v_\text{rel}^2+c_\text{s}^2$). The parameter $\xi$ depends on the geometry of the flow. In our modeling, we assume $\xi=1$ for simplicity. Finally, $\rho=n m_\text{p}$ is the density of the surrounding medium, which is the ISM in our case. 

 The Bondi formula is generally assumed to be an order-of-magnitude estimate of the upper limit of the accretion rate. It applies broadly however to different geometries of the accretion flow; for example, it can be used in the case of wind-accretion in wide binary systems. However, close to the compact object, an accretion disc can be formed.
 There is also the possibility that an accretion disc forms around an AINS, which forms the main study of this paper.

  Due to turbulence, the accreted matter at every moment has non-zero angular momentum. 
 Its amount can be estimated as follows \citep{2002A&A...381.1000P}.
 Turbulent vortices exist in a wide range of scales. However, for accretion, there is an important characteristic scale defined by the Bondi radius. Numerous vortices of smaller scales have different orientations, and their torques on the NS are averaged. Oppositely, a very large vortex with a scale much larger than $R_\mathrm{G}$ cannot produce a torque on an NS, because only a small fraction of the matter connected with this vortex can be captured and accreted. Thus, the spin behavior of the NS is mainly influenced by vortices of size $\approx R_\mathrm{G}$. 

 Let us estimate the specific angular momentum of matter in a vortex with the size $R_\mathrm{G}$,
 \begin{equation}
     j_\text{t}= v_\text{t}(R_\mathrm{G}) R_\mathrm{G},
 \end{equation}
where $v_\text{t}$ is the turbulent velocity. For low-velocity NSs, we have $R_\mathrm{G}\sim$~0.1-10~AU. At this scale, there are no robust measurements of the turbulence properties in the ISM. Measurements are available only at the scale $\gtrsim 100$~AU, see e.g., Figure~12 in \cite{2016SAAS...43...85K}.

Generally, it is expected that $v_\text{t}(r)\propto r^\alpha $, where $1/3 \lesssim \alpha \lesssim 1/2$. Here, the 1/3 limit corresponds to the Kolmogorov scaling, and 1/2 to magnetohydrodynamic (MHD) turbulence \citep{2021PASP..133j2001B}.
Observations over lengthscales of $\sim 0.01-100$~pc favour $\alpha\approx 1/2$ \citep{2014NPGeo..21..587F, 2016SAAS...43...85K}. At smaller scales, the situation becomes less certain. Thus, we proceed as follows. We perform calculations for an `optimistic' scenario (larger turbulent angular momentum available for an AINS), and then use several sets of parameters that result in a smaller turbulent velocity at small scales. 

 In the optimistic scenario, we assume $v_\text{t}(r)=v_\text{t}(R_t) (r/R_t)^{1/3}$, where $R_\text{t}=2\times 10^{20}$~cm and $v_\text{t}(R_t)=10$~km~s$^{-1}$. The most pessimistic scenarios with $\alpha= 1/2$ produce $v_\text{t}(R_\mathrm{G}$) up to one order of magnitude smaller. Note that at smaller scales the slope can become flatter \citep{2014NPGeo..21..587F}. 
 This might favour our optimistic scenario.  Relatively large values of the turbulent velocity at the scale $\lesssim 100$~AU are also supported by pulsar measurements \citet{2025ApJS..278...13L}.
 
 In the optimistic scenario, the maximum specific torque can be estimated as follows. The maximum value of the Bondi radius corresponds to $v=10$~km~s$^{-1}$: $R_\text{G}^\mathrm{max}=4\times 10^{14} \,v_{6}^{-2}$~cm, where $v_6 \equiv v/10^6\text{cm~s}^{-1}$.  The corresponding velocity is $v_\text{t}(R_\text{G}^\mathrm{max})=0.123$~km~s$^{-1}$, the specific turbulent torque is $j_\text{t}^\mathrm{max}=5\times10^{18}$~cm$^{2}$~s$^{-1}$. The accretion rate is $\dot M^\mathrm{max}=\pi (R_\text{G}^\mathrm{max})^2 n\, m_\text{p}\, v = 8\times 10^{11} n_{1} v_{6}^{-3} $~g~s$^{-1}$, where $n_1 \equiv n/1\text{~cm}^{-3}$. Thus, $J_\text{t}^\mathrm{max}=\dot{M}^\mathrm{max}j_\text{t}^\mathrm{max}\approx 4\times 10^{30}$~g~cm$^{2}$~s$^{-2}$. 


  The captured matter can form a disc around the accretor if its specific angular momentum is larger than the Keplerian momentum at the inner boundary of the flow. In the case of AINSs, the latter one can be well approximated by the magnetospheric radius. We assume that the magnetospheric radius for an \emph{accreting} NS is equal to the Alfv{\'e}n radius:

  \begin{equation} \label{eq:alfvenradius}
  \begin{aligned}
      R_\text{A}&=\left( \frac{\mu^2}{2\dot M\sqrt{2GM}} \right)^{2/7}\\&\approx8.2\times 10^9 \mu_{30}^{4/7} 
      \left( \frac{\dot M}{8\times 10^{11} \mathrm{~g~s}^{-1}}\right)^{-2/7} \, \mathrm{cm},
      \end{aligned}
  \end{equation}
  where the magnetic moment $\mu_{30} = \mu/(10^{30}\text{~G~cm}^3)$ and we have assumed a 1.4 solar mass neutron-star. Thus, $j_\text{K}=v_K(R_\mathrm{A}) R_\mathrm{A}$ and if $j_\text{t}>j_\text{K}$ then a disc is formed. 
  Note that, in general, a certain amount of toroidal field will be wound up as the dipole lines twist within the circling plasma which will necessarily change the local nature of pressure balance and one generally expects $R_{\rm A}$ to carry a prefactor (of order between $\sim$~0.1 and unity) depending on the rotation profile and microphysical aspects of the disc material \cite[see, e.g.,][]{gs21}.

 In any case, we see that $j_\text{t}\propto v^{-8/3}$, while $j_\text{K}\propto v^{3/7}$. Thus, $j_\text{t}/j_\text{K}\propto v^{-65/21}$. That is, the condition $j_\text{t}>j_\text{K}$ is valid only for the lowest-velocity AINSs with the largest accretion rate (and hence luminosity). As we will see, only small (e.g., decaying) magnetic fields, as $R_\text{A}\propto \mu^{4/7}$, can allow for a disc formation around AINSs with velocities of about a few tens of km~s$^{-1}$. In all cases, the circularisation radius defined by $j_\text{t}=\sqrt{GMR_\mathrm{cir}}$ is close to $R_\text{A}$. 

As the NS moves through the ISM, the external torque might change on a characteristic time scale $\tau_\text{t}\sim R_\text{G}/v\approx2GM/v^3\approx 12 \, v_6^{-3} $~yr. 

 Due to a low accretion rate, it is expected that AINSs are in the regime of settling accretion.
 This regime is characterised by slow cooling \citep{2012MNRAS.420..216S}. Thus, the accretion rate onto the surface is sufficiently reduced. The case of AINSs was analyzed by \cite {2015MNRAS.447.2817P}. 
 Disc formation requires that the matter can cool down. This poses a problem, as we need $t_\mathrm{cool}<\tau_\text{t}$. However, \cite{2015MNRAS.447.2817P} demonstrated that the cooling time can be of the same order as $\tau_\text{t}$ for fiducial parameters of an AINS. This makes the formation of a standard thin accretion disc problematic.
 In addition, at the settling accretion stage, part of the captured angular momentum can be carried away by convection, preventing the formation of an accretion disc \citep{2012MNRAS.420..216S}.
 Detailed analysis of disc formation around AINSs requires a realistic 3D modeling, which is beyond the scope of this paper. Still, for NSs with the lowest velocity accreting from relatively dense ISM, disc formation is quite realistic. Below, we assume that the disc can be formed if $j_\text{t}>j_\text{K}$. 

\subsection{Extrinsic torques}


A magnetised NS accreting matter with non-zero specific angular momentum would experience spin-up (denoted with an su subscript) and spin-down (sd) torques,
\begin{equation}
\label{eq:Euler}
    I\frac{\mathrm{d}\omega}{\mathrm{d} t}= K_\mathrm{su}-K_\mathrm{sd},
\end{equation}
where $\omega=2\pi/P$ is the spin frequency. 

For disc accretion, the maximum spin-up torque (when the disc is in the equatorial plane and the direction of its rotation coincides with the spin of the NS) can be calculated as:
\begin{equation}
    K_\mathrm{su}=\dot M \sqrt{GMR_\text{d}}.
    \label{eq:ksu_d}
\end{equation}
Here, $R_\text{d}$ is the inner disc radius. 
In the case of the disc accretion from the ISM, the relation between $R_\text{d}$ and $R_\text{A}$ might be different compared to `standard' models as the disc is not as massive and geometrically thin as in typical accreting X-ray binary systems, so we use $R_\text{d}=R_\text{A}$ in our calculations for simplicity.
Note that in our case, the spin-up torque is limited by the angular momentum available from the turbulised ISM: $J_\text{t}=\dot M j_\text{t} \lesssim 4\times 10^{30}$~g~cm$^{2}$~s$^{-2}$. 

The magnetic spin-down torque can be parameterised as
\begin{equation}
    K_\mathrm{sd}=\eta \frac{\mu^2}{R_\mathrm{co}^3},
    \label{eq:k_sd}
\end{equation}
where $R_\mathrm{co}=(GM/\omega^2)^{1/3}$ is the corotation radius \citep{1992ans..book.....L}. 
It is often assumed that $\eta \lesssim 1$, though in general will be sensitive to the magnetospheric conditions, twists, and general-relativistic effects.
 In addition, as the orientation of turbulent vortices is random, the disc can also contribute to the spin-down with the maximum torque defined by eq.~(\ref{eq:ksu_d}).
 If a disc forms, the situation can be more complicated for AINSs that rotate very slowly.
 If $R_\text{co}\gg R_\text{A}$ then the NS can be additionally spun up due to the interaction between the disc and the magnetic field lines in the region $R_\text{A}<r<R_\text{co}$.  However, we neglect this effect below. 

 \cite{2002A&A...381.1000P} demonstrated that in general, old AINSs might be very slowly rotating objects with periods of $\gtrsim$~months or even more. However, the transition to the stage of accretion happens at a critical period $P_A$, which can be about a few hundred seconds in the most optimistic case.  

Let us now estimate how rapidly AINSs are spinning down.  Using eq.~(\ref{eq:k_sd}) we obtain $K_\mathrm{sd}= 2 \eta \times 10^{27} \mu_{30}^2\, P_4^{-2}$~g~cm$^{2}$~s$^{-2}$. 
 Then, from $I \dot \omega = -K_\mathrm{sd}$ we obtain $\dot p _\mathrm{sd, \mu} \approx 3\times 10^{-11} \eta \mu_{30}^2 I_{45}^{-1}.$ Note, that this value is not spin-dependent. 

 As it was mentioned already, an additional spin-down can occur due to the accretion of matter with non-zero specific angular momentum. For an estimate, we can use eq.~(\ref{eq:ksu_d}). However, for the upper limit, we can apply $J_\text{t}^\mathrm{max}$. Then we obtain $\dot p_\mathrm{sd, max}= J_\text{t}^\mathrm{max} P^2/(2\pi I)= 10^{-8} J_{t, 30}^\mathrm{max} P_4^2 I_{45}^{-1}$. At any given moment, this spin-down can be larger than  $\dot p_\mathrm{sd, \mu} $. However, we are interested in the evolution on the time scale $P/\dot P$, which is much larger than $R_\text{G}/v$. So, the disc can contribute to spin-up as well as to spin-down. That is why we neglect this spin-down in the following estimate for the early stage of evolution of an AINS. Then we obtain that an AINS can have the spin period $\lesssim 10^4$~s for $\approx 10 \eta^{-1}$~Myr. An isolated NS starts accreting not earlier than when the critical period $P_\text{A}$, defined by the condition $R_\text{A}=R_\text{c}$, is reached:
 \begin{equation}
P_\text{A} \approx 320 \,\left(\frac{B}{10^{12}\text{~G}}\right)^{6/7}  \left(\frac{\dot{M}}{10^{11}\text{~g~s}^{-1}}\right)^{-3/7} \text{~s}.
    \label{eq:pa}
\end{equation}

 However, in our population synthesis (see below), we apply another condition for the transition from the propeller to the accretor stage: $R_\text{m}=R_\text{c}$ \citep{1981MNRAS.196..209D}. The magnetospheric radius at the propeller stage, $R_\text{m},$ is defined in Table~1. Then, the critical period is:
\begin{equation}
\begin{split}
P_\text{PA} \approx 2.6 \times10^4\left(\frac{v}{10\text{~\kms}}\right)^{-2/3}\left(\frac{B}{10^{12}\text{~G}}\right)^{2/3} \left(\frac{\dot{M}}{10^{11}\text{~g~s}^{-1}}\right)^{-1/3} \text{s}.
    \label{eq:pa_rm}
\end{split}
\end{equation}
 
 As the spin period of an AINS becomes longer, the spin-down due to magnetic braking ($K_\text{sd}$, eq.~\ref{eq:k_sd}) becomes too low, and the spin evolution is dominated by the external angular momentum from the turbulised ISM. Thus, finally, a kind of equilibrium is reached \citep{2002A&A...381.1000P}; however, the spin period of the AINS has large fluctuations around this quasi-equilibrium value, and on a very long time scale, the NS continues to slow down due to magnetic braking, eq.~(\ref{eq:k_sd}). 






\section{Population synthesis of isolated neutron stars}
\label{sec:pops}

In this study, we focus on AINSs. Thus, in the previous section we described the key features of these objects and our model assumptions. However, to obtain robust numbers of such sources in the Galaxy accounting for their lifetime, it is necessary to perform a population synthesis which includes kinematic evolution of NSs, properties of the ISM, and all evolutionary stages prior to the onset of accretion. In this section, we summarise the main properties of our population synthesis model. For details, we refer to \cite{2026JHEAp..5300643A}.
To begin, we discuss the population synthesis model for Galactic isolated NSs. 
Then, we describe new features implemented to calculate the properties of AINSs with discs.

Our population synthesis calculations can be divided into four steps. (i) Set the initial distributions of the NS parameters. Then, (ii) calculate the kinematic properties of NSs, i.e., their position and velocity in the Galaxy over the lifetime of the Galaxy $\sim13.6$~Gyr. Next, (iii) derive the evolutionary track, i.e., model the NS parameters over time, using their kinematic properties and the number-density map of the ISM. The evolutionary track contains information about the position and observable properties (flux, temperature) of AINSs. Finally, (iv) sum up the evolutionary tracks of all AINSs and apply a normalisation procedure to derive the properties of the global population. In this study, these properties, in the first place, include the percentage of dAINSs in the Galaxy and their parameters.

\subsection{Spin evolution of a neutron star}
\label{sec:evo}

Let us begin the description of the population calculations with the spin evolution of an NS. We consider the spin evolution following the general description \cite[see, e.g.,][]{1992ans..book.....L, 2024Galax..12....7A}. We fix the NS mass  as $M=1.4\,M_\odot$ and the moment of inertia $I=10^{45}$~g~cm$^2$ throughout. At the start of the evolution, the NS is assigned initial values of the spin period $P$ (or the spin frequency $\omega=2\pi/P$), the dipole surface magnetic field $B$ (or the magnetic moment $\mu = {\Bs}R^3$, $R=10$~km), the characteristic velocity $v$, and the accretion rate $\dot{M}$. The evolution of the spin period $P$ depends on the parameters ${\Bs}$, $v$, $\dot{M}$, the balance between the characteristic radii, listed in Table~\ref{tab:radii}, and the evolutionary stage between which the neutron star can transition.

\begin{center}
\begin{table*}
\centering
\caption{Characteristic radii used in the calculations of the NS evolution. 
\label{tab:radii}}
\renewcommand{\arraystretch}{1.8}
\begin{tabularx}{\textwidth}{ c @{\extracolsep{\fill}} p{7cm} @{\extracolsep{\fill}} p{8.4cm}}
\bottomrule
Symbol & Name & Expression \\
\midrule
$R_\text{G}$ & Gravitational capture (Bondi) radius & $R_\text{G}={2GM}/{v^2}$ \\
$R_\text{l}$ & Light cylinder radius & $R_\text{l} = {c}/{\omega}$ \\
$R_\text{co}$ & Corotation radius &  $R_{\text{co}} = \left( \frac{GM}{\omega^2} \right)^{1/3}$\\
$R_\text{c}$ & Radius of the centrifugal barrier & $R_{\text{c}} = 0.87 R_\text{co}= 0.87\left( \frac{GM}{\omega^2} \right)^{1/3}$\\
$R_\text{A}$ & Alfv{\'e}n radius (magnetosphere radius of accreting NSs) & $R_A = \left( \frac{ \mu^2}{2 \dot{M} \sqrt{2GM}} \right)^{2/7}$\\
$R_\text{m}$ & Magnetosphere radius at the propeller stage & $R_\text{m} = R_\text{A}^{7/9}R_\text{G}^{2/9} = \left(\frac{\mu^2 \sqrt{2GM}}{2\dot{M}v^2}\right)^{2/9}$\\
$R_\text{Sh}$ & Shvartsman radius & $R_{\text{Sh}}=\left( \frac{8 \mu^2 (GM)^2 \omega^4}{\dot{M} v^5 c^4} \right)^{1/2} = R_\text{G} \left(\frac{2 \mu^2 \omega^4}{\dot{M} v c^4}\right)^{1/2}$\\
$R_\text{Sh}^\text{env}$ & Shvartsman radius in the envelope & $R_\text{Sh}^\text{env} = \left(\frac{2\mu^2\omega^4\sqrt{2G M}}{\dot{M}v^2 c^4}\right)^{2} = R_\text{G}\left(\frac{2\mu^2\omega^4}{\dot{M}vc^4}\right)^2$\\ 
$R_\text{circ}$ & Circularisation radius & $R_\text{circ}=\frac{j^2_\text{t}}{GM} =\frac{4GM}{v^4}\left(v_\text{t}(R_\text{t}){R_\text{G}^{\alpha}}/{R_\text{t}^{\alpha}}\right)^2$ \\ 
\addlinespace[4pt]
\bottomrule
\end{tabularx}
\end{table*}
\end{center}



There are four main evolutionary stages of an INS: ejector, propeller, accretor and georotator.

INSs are usually born at the ejector stage. They produce a wind that prevents external matter from entering their magnetosphere. The wind power decreases as the rotation slows down. Eventually, the external matter pressure prevails, and the transition to the propeller stage occurs. At this stage, the magnetosphere interacts with the external material while the centrifugal barrier restricts this matter from reaching the NS surface. This interaction causes the NS to lose angular momentum further until the accretor stage begins. Alternatively, if the magnetosphere is larger than the gravitational capture radius, an ``exotic'' georotator stage can occur.

Throughout these stages, an INS can change its rotational energy at different rates and due to different mechanisms. This results in different expressions for the spin-down and spin-up torques in the Eq.~\ref{eq:Euler}. 
The spin-down torques are listed in Table~\ref{tab:torque}. At the ejector stage, rotational energy is lost due to the pulsar wind emission. At other stages, the spin period evolves due to interactions between the magnetosphere and external matter. For the propeller stage, four models with different mechanisms of rotational energy loss are considered. They are listed in order of the decreasing spin-down rate: model A \citep{1975SvAL....1..223S}, B \citep{1973ApJ...179..585D}, C \citep{1975AA....39..185I}, and D \citep{1981MNRAS.196..209D}. 

In the calculations, the spin-up torque at each stage is $K_\text{su}=0$.  Here we simplify the spin evolution at the accretor stage -- we neglect the stochastic torque of the accreted material. This is done for two reasons. First, a detailed calculation of the spin period of the accreting NS is not required, because after the transition to the accretor stage, the period increases rapidly (on a timescale $<1$~Gyr), and thereafter it is used only to check whether the reverse accretor–propeller transition occurs. Second, as shown in Section~\ref{sec:acc}, the torque changes sign on a time scale of $\sim12v_6^{-3}$~yr, while integrating over $13.6$~Gyr requires a much larger timestep. Accounting for it would require averaging over many sign changes. Since the torque is stochastic with no preferred sign, the overall effect is small and does not systematically bias the evolution.

\begin{center}
\begin{table}
\centering
\caption{Spin-down torques for evolutionary stages of an NS. For the propeller stage, four spin-down models are considered. Here, the free-fall velocity is $v_\text{ff}(R_\text{m})=\sqrt{2GM/R_\text{m}}$.
\label{tab:torque}}
\renewcommand{\arraystretch}{1.2}
\tabcolsep=0pt
\begin{tabular*}{20pc}{@{\extracolsep\fill}l p{4cm}}
\bottomrule
\addlinespace[4pt]
Stage & Spin-down torque \\
\midrule
Ejector & $K_\text{E} = 2\mu^2\omega^3/c^3 = 2 {\mu^2}/{R_\mathrm{l}^3}$  \\ 
Propeller (model A) & $K_\text{P} = \dot{M} \omega R_{\text{m}}^2$  \\
Propeller (model B) & $K_\text{P} = \dot{M} \sqrt{2GMR_{\text{m}}}$  \\
Propeller (model C) & $K_\text{P} = \dot{M}\, \text{max}[v^2,~v^2_{\text{ff}}(R_\text{m})]$  \\
Propeller (model D) & $K_\text{P} = \dot{M} v^2 / (2\omega)$  \\
Accretor & $K_{\text{A}}=0.4{\mu^2}/{R_{\text{c}}^3}$  \\
Georotator & $K_\text{G} = 0$  \\
\bottomrule
\end{tabular*}
\end{table}
\end{center}

Under the considered approach, the transition conditions are expressed as equations in terms of characteristic radii. These conditions are listed in Table~\ref{tab:transition}. 
 They reflect changes in modes of interaction with external matter and define the critical spin period for transitions between evolutionary stages. 
 Note that for accretor-propeller (direct and reverse) transitions, we use the centrifugal barrier radius introduced by \cite{2023MNRAS.520.4315L} instead of the corotation radius $R_\text{co}$ used in many studies. 

\begin{center}
\begin{table}
\centering
\caption{Transitions between evolutionary stages of an NS expressed in terms of characteristic radii. 
\label{tab:transition}}
\renewcommand{\arraystretch}{1.2}
\tabcolsep=0pt
\begin{tabular*}{20pc}{@{\extracolsep\fill}lp{5cm}}
\bottomrule
\addlinespace[4pt]
\multicolumn{2}{c}{Direct transition condition} \\
\midrule
Ejector-Propeller & $R_{\text{Sh}} \le \text{max}(R_\text{G},~R_\text{l})$ \\ 
 & \hspace{0.9cm} and $R_{\text{Sh}}^\text{env} \le \text{min}(R_\text{G},~R_\text{l})$ \\ 
Propeller-Accretor & $R_\text{m} \le R_\text{c}$  \\
Accretor-Georotator & $R_\text{A} \ge R_\text{G}$  \\
\bottomrule
\addlinespace[4pt]
\multicolumn{2}{c}{Reverse transition condition} \\
\midrule
Propeller-Ejector & $R_{\text{Sh}} > \text{max}(R_\text{G},~R_\text{l})$ and $R_\text{m} > R_\text{l}$ \\ 
Accretor-Propeller & $R_\text{m} > R_\text{c}$  \\
Georotator-Accretor & $R_\text{A} < R_\text{G}$  \\
\bottomrule
\end{tabular*}
\end{table}
\end{center}



For the evolution of the surface (dipolar) magnetic field, we consider two models: a constant field (CF), where $B$ is held at its initial value $B_0$, and an exponentially decaying field (ED). The magnetic field evolution in the latter case is as follows:
\begin{equation}
\Bs = {\Bs}_0 \exp\{-t/\tau\},
\end{equation}
where $t$ is the age of the NS and $\tau=1.5 \times 10^{9}$~yr is the decay timescale.

\subsection{Model of the Milky Way}

The model of the Milky Way includes the gravitational potential, the velocity and density maps of the ISM. These models are the same as those adopted in \cite{2026JHEAp..5300643A}. In this subsection, we describe the models only briefly and mostly qualitatively.

The gravitational potential consists of the \cite{1975PASJ...27..533M} disc, a \cite{1996ApJ...462..563N} halo, and a \cite{1990ApJ...356..359H} potential for both Galactic bulge and nucleus. 

Near the Galactic plane, the velocity of the ISM is equal to the circular velocity $\vec{v}_\text{circ}  = v_\text{circ} \{-{y}/R,~{x}/{R},~0\}$, where $v_\text{circ}$ is derived from the gravitational potential. In the Galactic halo, the medium rotates with a constant velocity $180$~\kms in the same direction as the disc \citep{2016ApJ...822...21H}. For continuity, between the disc and the halo the velocity transitions as follows
\begin{equation}
\label{eq:vISM}
v_\text{ISM} = v_\text{circ} - z\frac{\partial v}{ \partial z},
\end{equation}
where $\partial v/ \partial z = 15$~\kms~kpc$^{-1}$ \citep{Marasco2011-vd}.

For the number density map $n(R,~z)$, we adopt two models: a simple (or one-phase) and a two-phase model. The simple model considers only the cold medium, which has  the speed of sound of $c_\text{s}=10$~\kms  and includes the molecular, cold neutral, and warm ionized medium. The total number density map is adopted as the sum of the distributions proposed by \cite{2006A&A...459..113M, 2017ApJ...835...29Y, 2008PASA...25..184G}.

In the two-phase model, both cold and hot phases are present. The hot phase is several orders of magnitude less dense than the cold one. It has $c_\text{s}=100$~\kms and includes the coronal gas in the Galactic plane and the hot gas from the Galactic halo. An NS at any point on the trajectory can be in either phase. The distribution of the hot phase is adopted from \cite{2024A&A...681A..78L}, while the cold phase distribution is now governed by the pressure balance between the two phases. Thus, near the Galactic plane, there is only the cold phase, while higher in the Galactic halo, an NS is almost always in the hot phase. The mean number density distribution of the cold phase is kept the same as in the simple model. 
In the two-phase model, along the trajectory, the NS experiences much larger fluctuations in the number density of the ISM than in the one-phase model.

The trajectory and the velocity vector of the NS $\vec{v}_\text{NS}$ are calculated in the inertial reference frame using the adopted Galactic potential. Then, the parameter $v$ that is used in the evolutionary calculations is defined as 
\begin{equation}
    v = \sqrt{v_\text{rel}^2+c_\text{s}^2} =\sqrt{|\vec{v}_\text{ISM} - \vec{v}_\text{NS}|^2 + c_\text{s}^2}.
\end{equation}
Using the number density map (one- or two-phase model), we calculate the accretion rate along the trajectory as $\dot{M} = \pi R_\text{G}^2 n m_\text{p}v$, where $m_\text{p}$ is the proton mass.

\subsection{Initial population}

To start the calculations, it is necessary to specify the initial distributions for the NS population. It includes the position, velocity, spin period, and magnetic field. Here we list the probability density functions of these parameters.

Following \cite{2004A&A...422..545Y}, the distance between the NS progenitors and the center of the Milky way in the Galactic plane is assumed to be distributed as
\begin{equation}
\label{eq:f_r}
    f (R)\propto \left(\frac{R}{R_\odot}\right)^a \exp \left[-b\left(\frac{R}{R_\odot}\right)\right],
\end{equation}
where $R_\odot=8$~kpc is the galactocentric distance of the Sun, $R=\sqrt{x^2+y^2}$, $a = 4$, $b = 6.8$. 

The height above the Galactic plane is the vertical coordinate $z$. 
Its absolute value is distributed as 
\begin{equation}
    f(|z|) \propto \exp \left(\frac{|z|}{z_0}\right),
    \label{eq:f_z}
\end{equation}
where $z$ is the vertical distance from the Galactic plane, $z_0 = 50$~pc \citep{2006ApJ...643..332F}. The coordinate $z$ is $<0$ in half the cases.

The velocity vector of a newborn NS is calculated as a sum of three components $(\vec{v}_\text{NS})_0 = \vec{v}_\text{circ} + \vec{v}_\text{res} + \vec{v}_\text{kick}$. Here $\vec{v}_\text{circ}$ is the circular velocity in the Galactic potential. The direction of the residual velocity $\vec{v}_\text{res}$ of the progenitor star is uniform over a sphere and the components $((v_\text{res})_x,(v_\text{res})_y,(v_\text{res})_z)$ undergo the normal distribution with standard deviations $(10,10,8)$~\kms \citep{2022AstL...48..243B}. Finally, $\vec{v}_\text{kick}$ is the kick velocity gained after the supernova explosion. Its direction is uniform; the distribution of the absolute value is the weighted sum of two Maxwellian distributions \citep{2021MNRAS.508.3345I}: $f(v) = wf_{\sigma1}(v) + (1-w)f_{\sigma2}(v)$, where $w=0.2$, $\sigma_1=45$~km~s$^{-1}$, $\sigma_2=336$~km~s$^{-1}$.

In our modeling, depending on the initial parameters, we divide the NSs into `pulsars' and `magnetars'. 
Roughly speaking, we expect that the Galactic NS population consists of $90$\% pulsars and $10$\% magnetars \citep{1999PNAS...96.5351K}, which we use as a basis to set the initial field strengths. In our modeling, they differ only by the initial distributions of the spin period and the magnetic field rather than by some observational definition related to high-energy activity. The logarithm of each parameter $x_0$ (either $P_0$ or ${\Bs}_0$) is assumed to be distributed normally,
\begin{equation}
f (\log_{10} x_0) = \frac{1}{\sigma_{x_0} \sqrt{2 \pi}} \exp \left( {- \frac{(\log_{10} x_0 - \overline{\log_{10} x_0})^2}{2\sigma_{x_0}^2}} \right).
\end{equation}

For the pulsar distribution, the initial spin period (in seconds) is set through $\overline{\log_{10} P_0}= -1.04$, $\sigma_{P_0} = 0.53$. The initial magnetic field (in Gauss) from $\overline{\log_{10} \Bs}= 12.44$, $\sigma_{\Bs} = 0.44$ \citep{2022MNRAS.514.4606I}.

While the evolution over $13.6$~Gyr of a NS born as a pulsar can be estimated more simply through equations given in Section~\ref{sec:evo}, the evolution of the magnetar population requires an additional, short step before the main calculations. This additional step characterises the rapid decay of the magnetar magnetic field over the first several million years, which is negligible in comparison to $13.6$~Gyr. It is only supposed to yield initial parameters (spin period and dipole magnetic field), so other data is not saved. 

The initial parameters of magnetars are calculated as follows. At first, magnetars have the same initial spin period distribution as pulsars. 
The magnetic field is taken from the distribution with parameters $\overline{\log_{10}{\Bs}_0'}=14.33$, $\sigma_{{\Bs}'_0} = 0.46$ \citep{2014ApJS..212....6O}. Then, their spin period evolves as at the ejector stage, while their dipolar magnetic field decays as follows:
\begin{equation}
\label{eq:B_short}
    {\Bs}(t)={\Bs}'_0\frac{\exp(-t/\tau_{\text{Ohm}})}{1+(\tau_{\text{Ohm}}/\tau_{\text{Hall}})(1-\exp(-t/\tau_{\text{Ohm}}))},
\end{equation}
where we fix $\tau_{\text{Ohm}} = 10^6$~yr and $\tau_{\text{Hall}} = 10^4/({\Bs}_0/10^{15}\text{~G})$~yr for concreteness \citep{2008ApJ...673L.167A}. If the field is ultra-strong ($\gtrsim 10^{14}$~G) then plastic flow may also adjust the magnetic evolution, but we ignore such complications here. This additional step of the evolution is calculated until the magnetic field drops to $\sim 20$~times lower than the initial value, which is $\sim 3$ $e$-foldings \citep{2014MNRAS.438.1618G}. Then, the spin period and magnetic field reached at the end of the rapid decay stage are stored as the initial spin period and magnetic field values for the evolution over $13.6$~Gyr, which is similar for both pulsars and magnetars, and is described in Section~\ref{sec:evo}.

\subsection{Calculation of the number of NSs with accretion discs}

Here, we describe our method for calculating the number of NSs with accretion discs and their parameter distributions. We perform calculations for the turbulent velocity distribution with the Kolmogorov scaling $v_\text{t}(r)\propto r^{1/3}$. At the end of this section, we also briefly discuss the results for the MHD turbulence where $v_\text{t}(r)\propto r^{1/2}$.

We generate the distribution of all the necessary parameters and draw the initial coordinates and spatial velocity to calculate $N_\text{tr} = 10^7$ distinct trajectories using a supercomputer cluster. Then we check whether the NS remains in the Galaxy, before storing the coordinates and characteristic velocity $v$ for each track. We save trajectories only for the objects that always stay within $100$~kpc from the Galactic center.
They constitute $\approx52.6\%$ of all considered sets of parameters. 
Thus, we define the fraction $f_\text{b}\approx0.526$ of the calculated NSs remaining bound to the Galaxy.

Each of $f_\text{b}\times N_\text{tr}\approx5.26$~million trajectories are used $16$ times for spin evolution calculations within two models of the ISM, two models of the magnetic field evolution, and four models of the torque at the propeller stage. {The time step is chosen to be $10$~Myr, about twenty times shorter than the Galactic orbital period of the nearby NSs, $\sim200$~Myr, so the external conditions change only slightly between steps. Thus, each track has $1361$ time points.} Once the evolution of an NS has been calculated, we use the parameters $v$, $\dot{M}$, and ${\Bs}$ to determine the ratio of the turbulent to Keplerian specific torques,
\begin{equation}
\label{n_kol}
    \begin{aligned}
    n_\text{d} &= \frac{j_\text{t}}{j_\text{K}} \equiv \sqrt{\frac{R_\text{circ}}{R_\text{A}}} \\ 
    &\approx 3.9~ \left(\frac{v}{10\text{~\kms}}\right)^{-65/21} \left(\frac{{\Bs}}{10^{12}\text{~G}}\right)^{-2/7} \left(\frac{n}{1\text{~cm}^{-3}}\right)^{1/7},
\end{aligned}
\end{equation}
along the trajectory, assuming that the turbulent velocity is described by the Kolmogorov scaling. We assume that if the NS is at the accretor stage and $n_\text{d}>1$, an accretion disc 
is formed around the NS magnetosphere. This condition corresponds to the circularisation radius being greater than the Alfv{\'e}n radius ($R_\text{circ}>R_\text{A}$). 
{The condition for disc formation depends only on quantities that are considered constant over $10$~Myr: the NS velocity relative to the ISM, the magnetic field, and the ISM number density. Thus, if the condition is fulfilled at a given time step, the disc is assumed to exist during the following $10$~Myr. The sign of the stochastic torque and the disc orientation may change many times within this interval due to the evolving turbulence at the Bondi scale (see Sec.~\ref{sec:acc}), but this does not affect the disc-formation condition.}\footnote{Still, if the disc sometimes disappears for a short period of time during our time step, then it results in a slight overestimation of the number of dAINSs in our model.}

When summarising all data, we assign a weight $w_i$ to each point on every track that is proportional to the star formation rate (SFR) at a lookback time equal to the age $t_i$ of the NS. We adopt the SFR function from \cite{2016A&A...589A..66H} (the 5th curve in their Figure~4). This SFR function yields an overabundance of the NSs with ages of $10-13$~Gyr, while there are almost no NSs with $t\approx7-8$~Gyr. For each track, we normalise the weights in the following way:
\begin{equation}
    \sum_{i=1}^{1361} w_i = 1.
\end{equation}
After normalisation, for the points of the track where the NS does not accrete or does not have an accretion disc ($n_\text{d}<1$), we set the weights $w_i = 0$. This effectively counts the fraction of time that an NS spends as an accretor with a disc. The number of the NSs with discs is
\begin{equation}
    N_{\text{D}} = \dfrac{N}{f_{\text{b}} N_{\text{tr}}} 
\sum_{f_{\text{b}} N_\text{tr}} \sum_{i=1}^{1361} w_{i},
\end{equation}
where $N=3\times10^8$ is the assumed total number of NSs in the Milky Way. 

To demonstrate the properties of NSs, we plot histograms of various parameters. When obtaining the histograms, we treat the parameter values at each $t_i$ as unique measurements, each with a corresponding weight $w_i$.

For error estimation, we split $10^7$ tracks into $95$ groups and calculate the $N_\text{D}$ with the corresponding normalisation for each subsample. The mean value and standard deviation are calculated using \verb|numpy.mean| and \verb|numpy.std|.

In addition to the number of the dAINSs, we calculate their observable properties: the effective temperature of the polar caps $T$ and the X-ray flux $F_\text{X}$ accounting for the interstellar absorption. The procedure for obtaining $T$ and $F_\text{X}$ is the same as used by \cite{2026JHEAp..5300643A}. Due to numerous uncertainties, we do not take into account the luminosity of the accretion disc. We use a simplified model, assuming that all radiation is emitted at the polar caps. The radius of the polar caps is $R_\text{cap} = R_\text{NS} \sqrt{{R_\text{NS}}/{R_\text{A}}}$ \citep{1998astro.ph..4047S}, and the effective temperature measured by a distant observer is
\begin{equation}
    \begin{split}
    T=T_\text{loc}\sqrt{1-r_\text{g}/R_\text{NS}}=\left(\frac{(L_\text{X})_\text{loc} }{S_\text{cap}\sigma_\text{B}}\right)^{1/4}\sqrt{1-r_\text{g}/R_\text{NS}}=
    \\=\left(\frac{L_\text{X} }{S_\text{cap}\sigma_\text{B}\sqrt{1-r_\text{g}/R_\text{NS}}}\right)^{1/4}\sqrt{1-r_\text{g}/R_\text{NS}},
\end{split}
\end{equation}
where $T_\text{loc}$ and  $(L_\text{X})_\text{loc}$ are the local temperature and local luminosity at (or close to) the NS surface, $L_\text{X} = {GM\dot{M}}/{R_\text{NS}}$, $S_\text{cap} = 2\pi R_\text{cap}^2$, $\sigma_\text{B}$ is the Stefan-Boltzmann constant, $r_\text{g}=2GM/c^2$, and the factor $\sqrt{1-r_\text{g}/R_\text{NS}}\approx0.8$. 

The flux without absorption measured by a distant observer is 
\begin{equation}
\label{eq_f_0}
   (F_\text{X})_0 = \frac{L_\text{X}}{4\pi d^2} \sqrt{1-r_\text{g}/R_\text{NS}}, 
\end{equation}
where $d$ is the distance between the NS with coordinates $(x,y,z)$ and the Sun at $(0,8,0)$~kpc. 

When calculating interstellar absorption, we consider the propagation of the blackbody spectrum $B_\nu(T)$ through the ISM, where the column density $N_\text{H}$ is calculated along the line of sight from the NS to the Sun. This is done using a density map of the molecular and neutral interstellar gas.

\begin{equation}
    F_\text{X}  =  (F_\text{X})_0\frac{\int_0^{\infty} \frac{B_\nu(T)}{h\nu} \exp({-\sigma(\nu) N_\text{H}}) \text{d}\nu}{\int_0^{\infty} {B_\nu(T)} \text{d}\nu}.
    \label{eq:flux}
\end{equation}

Here, the blackbody spectrum is $B_\nu(T) = {2h\nu^3}/[{c^2} ({\exp\{h\nu / k_\text{B}T\}-1})]$, where $h$ is the Planck constant and $k_\text{B}$ is the Boltzmann constant. It is important to note that interstellar extinction strongly depends on the photon energy distribution. Thus, the absorbed flux can be significantly different from eq.~(\ref{eq:flux}), if the spectrum deviates from the blackbody spectrum assumed in our simplified approach.

\section{Results of population calculations}
\label{sec:results}

Here, we present our results on the number of AINSs with discs and discuss basic properties of these sources.

\begin{figure*}
\includegraphics[width=\textwidth]{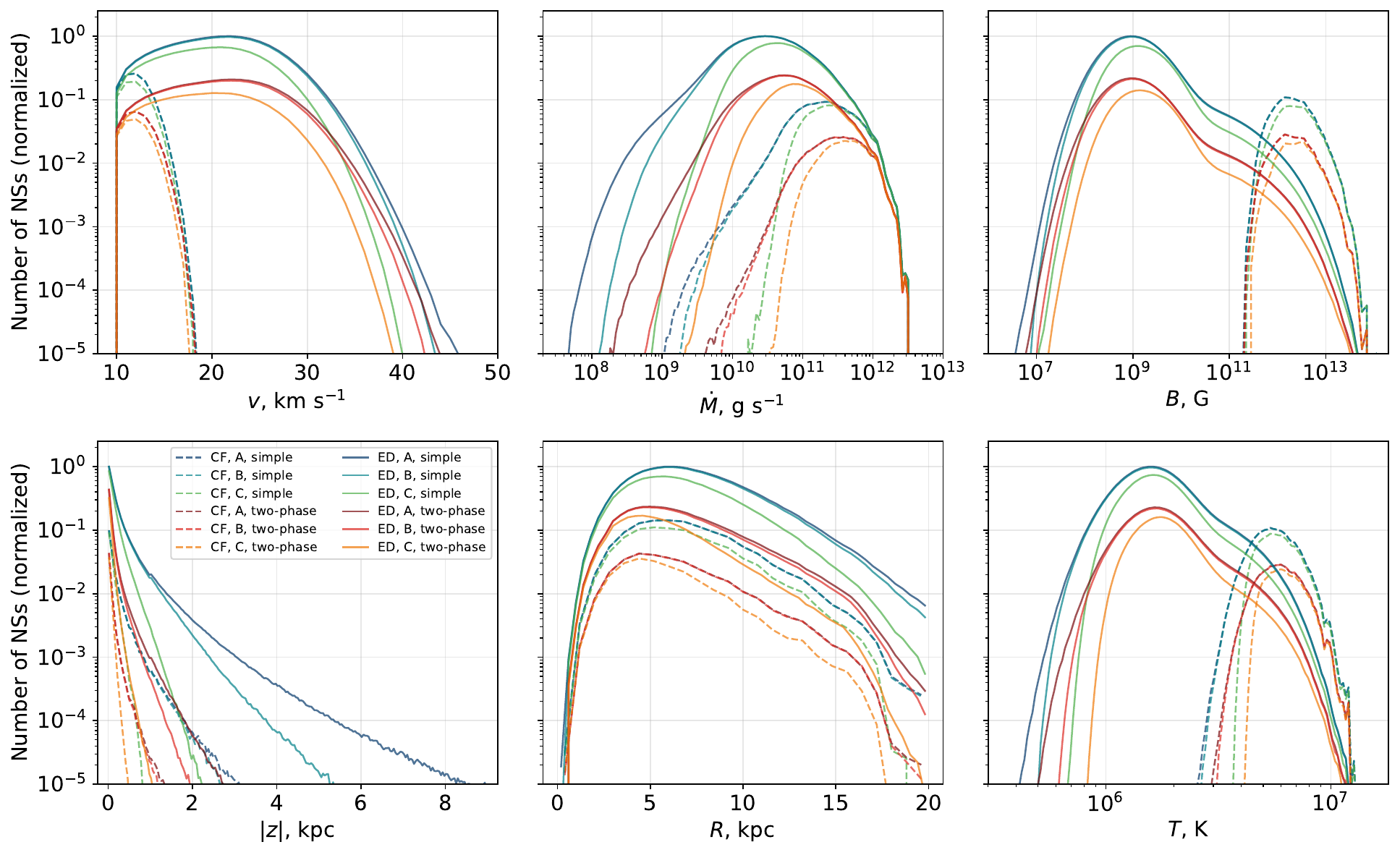}	
\centering 
\caption{Distribution of accreting NSs with discs over six parameters: characteristic velocity $v$, accretion rate $\dot{M}$, surface magnetic field $B$, height above the galactic plane $z$, the distance from the center of the Galaxy in the galactic plane $R$, and effective temperature of polar caps $T$. 
There are 12 models considered: two models of magnetic field evolution -- constant (CF) and decaying field (ED), three propeller models -- A, B, C, and two ISM models -- simple (one-phase) and two-phase model. All distributions are normalised so that the maximum value on each plot is 1. The legend is common to all panels.
}
\label{fig_six}
\end{figure*}

Table~\ref{tab:results} shows the calculated fractions of the NSs with accretion discs relative to the total population of NSs in the Milky Way. The number of accreting NSs here is taken from \cite{2026JHEAp..5300643A}, since the population synthesis design is the same. The dAINSs properties are shown in Figure~\ref{fig_six}. 

\begin{table*}
\centering
\caption{The number of accreting NSs (AINSs) and accretors with discs (dAINSs) in the Galaxy calculated in different models, assuming Kolmogorov turbulence in the ISM ($\alpha=1/3$).  Here $N_\text{D}$ is the number of accretors with discs, $N=3\times10^8$ is the total number of all NSs staying in the Galaxy, $N_\text{A}$ is the number of accretors. The propeller model D is shown only by the order of magnitude due to the high relative error. A zero value means that there are no dAINSs in the calculations within the given model.
}
\centering
\label{tab:results}
\renewcommand{\arraystretch}{1.1}
\begin{tabularx}{\textwidth}{@{\extracolsep{\fill}} l p{3.5cm} p{2.9cm} p{2.9cm} p{2.9cm}}
\bottomrule
\addlinespace[4pt]
\multicolumn{1}{l}{} & \multicolumn{4}{l}{One-phase ISM, constant field} \\
\midrule
Propeller model & $N_\text{D}$ & $N_\text{D}/N$, \% & $N_\text{A}/N$
\, \% & $N_\text{D}/N_\text{A}$, \% \\
A & $(4.3 \pm 0.4)\times10^5$ & $0.143 \pm 0.012$ & $44.7 \pm 0.4$ & $0.32 \pm 0.03$\\
B & $(4.3 \pm 0.4)\times10^5$ & $0.143 \pm 0.012$ & $32.5 \pm 0.5$ & $0.44 \pm 0.04$\\
C & $(3.16 \pm 0.26)\times10^5$ & $0.105 \pm 0.009$ & $1.61 \pm 0.15$ & $6.5 \pm 0.4$\\
D & $0$ & $0$ & $\lesssim10^{-4}$ & $0$ \\
\bottomrule
\addlinespace[4pt]
\multicolumn{1}{l}{} & \multicolumn{4}{l}{One-phase ISM, exponentially decaying field} \\
\midrule
Propeller model & $N_\text{D}$ & $N_\text{D}/N$, \% & $N_\text{A}/N$, \% & $N_\text{D}/N_\text{A}$, \% \\
A & $(6.28 \pm 0.17)\times10^6$ & $2.09 \pm 0.06$ & $52.8 \pm 0.7$ & $4.0 \pm 0.5$\\
B & $(6.06 \pm 0.14)\times10^6$ & $2.02 \pm 0.05$ & $27.6 \pm 0.6$  & $7.3 \pm 0.8$ \\
C & $(4.03 \pm 0.09)\times10^6$ & $1.34 \pm 0.03$ & $5.4 \pm 0.4$ & $24.9 \pm 2.3$\\
D & $0$ & $0$ & $\lesssim10^{-4}$ & $0$ \\
\bottomrule
\addlinespace[4pt]
\multicolumn{1}{l}{} & \multicolumn{4}{l}{Two-phase ISM, constant field} \\
\midrule
Propeller model & $N_\text{D}$ & $N_\text{D}/N$, \% & $N_\text{A}/N$, \% & $N_\text{D}/N_\text{A}$, \% \\
A & $(1.13 \pm 0.12)\times10^5$ & $0.038 \pm 0.005$ & $36.2 \pm 0.4$ & $0.104 \pm 0.009$\\
B & $(1.13 \pm 0.13)\times10^5$ & $0.038 \pm 0.005$ & $23.6 \pm 0.6$ & $0.159 \pm 0.017$\\
C & $(8.29 \pm 0.10)\times10^4$ & $0.028 \pm 0.004$ & $1.5 \pm 0.14$ & $1.84 \pm 0.10$\\
D & $\lesssim3\times10^2$ & $\lesssim10^{-4}$ & $\lesssim10^{-3}$ & $\lesssim100$ \\
\bottomrule
\addlinespace[4pt]
\multicolumn{1}{l}{} & \multicolumn{4}{l}{Two-phase ISM, exponentially decaying field} \\
\midrule
Propeller model & $N_\text{D}$ & $N_\text{D}/N$, \% & $N_\text{A}/N$, \% & $N_\text{D}/N_\text{A}$, \% \\
A & $(1.37 \pm 0.04)\times10^6$ & $0.458 \pm 0.013$ & $41.3 \pm 0.6$ & $1.11 \pm 0.12$\\
B & $(1.30 \pm 0.04)\times10^6$ & $0.435 \pm 0.012$ & $17.3 \pm 0.7$ & $2.5 \pm 0.3$\\
C & $(8.10 \pm 0.24)\times10^5$ & $0.270 \pm 0.008$ & $2.93 \pm 0.22$ & $9.2 \pm 0.7$\\
D & $\lesssim3\times10^3$ & $\lesssim10^{-3}$ & $\lesssim10^{-3}$ & $\lesssim100$ \\
\bottomrule
\end{tabularx}
\end{table*}

Generally, dAINSs account for up to $\sim2$\% of the total number of INSs in the Milky Way, which, in our normalisation, corresponds to $\sim6\times10^6$ objects in the Galaxy (within uncertainties, the number is $\sim$a few $\times 10^6$--$10^7$).

The number of dAINSs varies greatly depending on the chosen model, and can be close to zero in certain circumstances. For instance, the propeller model D yields $0$ or $\lesssim  10^3$ dAINS. Model D differs from other propeller models due to an extremely inefficient spin-down mechanism which allows an INS to reach the accretor stage only in the case of high magnetic field $>10^{14}$~G, low velocity $v\sim10$~\kms, and high number density of ISM $>10$~cm$^{-3}$ at the same time. Objects with these parameters are rare, therefore, there are only a few accretors, and consequently, only a few INSs have accretion discs. In the other propeller stage models (A, B, and C), the dAINSs percentage is reliably above zero and ranges from $0.03$\% to $2$\%, depending on the ISM and magnetic field models. Below, we discuss the results excluding model D. 

The main property determining whether the NSs would have an accretion disc is low velocity $v$, which is best illustrated by the $v$ distribution of dAINSs shown in Figure~\ref{fig_six}. NSs with accretion discs do not have characteristic velocities above $50$~\kms (ED model) or $20$~\kms (CF model). All of these NSs are located in the cold phase of the ISM. Although low velocity values are favourable for the onset of accretion, the NSs can begin to accrete with $v\sim100$~\kms or higher. Disc formation requires a much lower velocity than the propeller-accretor transition. This is due to the strong dependence of the torque ratio $n_\text{d}\propto v^{-65/21}$, while the dependence on the accretion rate and the magnetic field is not so strong -- AINSs can have different $\dot{M}$ and $B$ values and still have accretion discs. Also, these two parameters are not correlated with each other. Thus, the INSs with accretion discs originate from a low-velocity tail of the $v$ distribution of AINSs. 

The second most important factor for the accretion disc formation is the magnetic field. The dependence of the number of dAINSs on the magnetic field can be seen when comparing the CF and ED models. If the magnetic field of the NSs decays exponentially by the factor $\sim 10^4$ over $13.6$~Gyr (ED model), the number of dAINSs is more than an order of magnitude greater than in the case of a constant field. This is related to the dependence of the Keplerian specific torque on the magnetospheric radius at the accretor stage $j_\text{K}\propto R_\text{A}^{1/2}\propto {\Bs}^{2/7}$, so $n_\text{d}\propto {\Bs}^{-2/7}$. For disc formation, the turbulent specific torque must exceed the Keplerian one, and this is easier to achieve with a smaller magnetosphere. Figure~\ref{fig_six} illustrates this effect more clearly. The magnetic field distribution in the ED case has a prominent peak at $\sim10^9$~G, whereas in the CF model, there are almost no NSs with ${\Bs}\lesssim10^{11}$~G and the number of dAINSs is approximately an order of magnitude lower. Additionally, a low magnetic field allows the AINSs with higher velocity to have accretion discs, so the $v$ distribution is much wider in the model with an exponentially decaying field in comparison to the CF model.

The influence of the number density distribution of the ambient medium is less significant, but still visible. In the two cases with the same propeller and magnetic field models, the two-phase ISM model yields several times fewer dAINSs than in the case of the simple ISM. The comparison of the $\dot{M}$ distributions in Figure~\ref{fig_six} shows that the accretion rate in the simple ISM model is $\approx2$ times higher than in the two-phase model. Higher $\dot{M}$ favours earlier accretion onset and disc formation. 


Surprisingly, the choice of propeller model does not affect the results significantly if we consider models that can produce a non-negligible number of accretors in general. The number of accretors $N_\text{A}$ decreases several dozen times from propeller model A to C within the same ISM and magnetic field model. At the same time, the number of accretors with discs $N_\text{D}$ does not fluctuate that much -- $N_\text{D}$ differs between models A, B, and C at most by a factor of two. So, the fraction $N_\text{D}/N_\text{A}$ increases as the propeller torque decreases, i.e., from models A to C. This is because the conditions for disc formation are stricter than those for the onset of accretion, mostly due to the low velocity. Thus, most of the NSs that have accretion discs in the propeller model A have the parameters favourable to reach the accretor stage and acquire an accretion disc in models B and C, too.

Figure~\ref{fig_six} shows the spatial distribution of dAINSs, as a function of height above the Galactic plane $z$ and distance from the Galactic center within the Galactic plane $R$. The $R$ distribution is almost identical to that of all the NSs in the Galaxy. Therefore, an accreting NS can have an accretion disc regardless of its distance from the Galactic center. The dAINSs are located in a thin disc near the Galactic plane in all of the considered evolutionary models. The main reason is that disc formation requires low velocities relative to the ISM, and these objects can not rise high above the Galactic plane. Generally, the greater the number of dAINSs in the model, the more scattered they will be; therefore, model ED yields a wider distribution than model CF. Similarly, propeller models A and B lead to a wider $|z|$-distribution than model C.
Another factor influencing the spatial distribution is the ISM model. In the simple ISM model, $68.27$\% ($99.73$\%) of dAINSs are located within $80-240$~pc ($2-4$~kpc) from the Galactic plane, while in the two-phase ISM, $68.27$\% ($99.73$\%) of dAINSs are located within $30-80$~pc ($0.3-1.4$~kpc). This is mainly because, in the two-phase model, the higher the NS is above the plane, the less chance it has of entering the cold medium with a sound speed  $c_\text{s}=10$~\kms. Otherwise, it is located in the hot ISM phase with $c_\text{s}=100$~\kms, resulting in a characteristic velocity of $v>100$~\kms; therefore, the NS does not have the opportunity to form a disc. In contrast, in the one-phase model, the entire medium is cold, so an NS located slightly above the plane can still have an accretion disc.

\begin{figure}
	\includegraphics[width=\columnwidth]{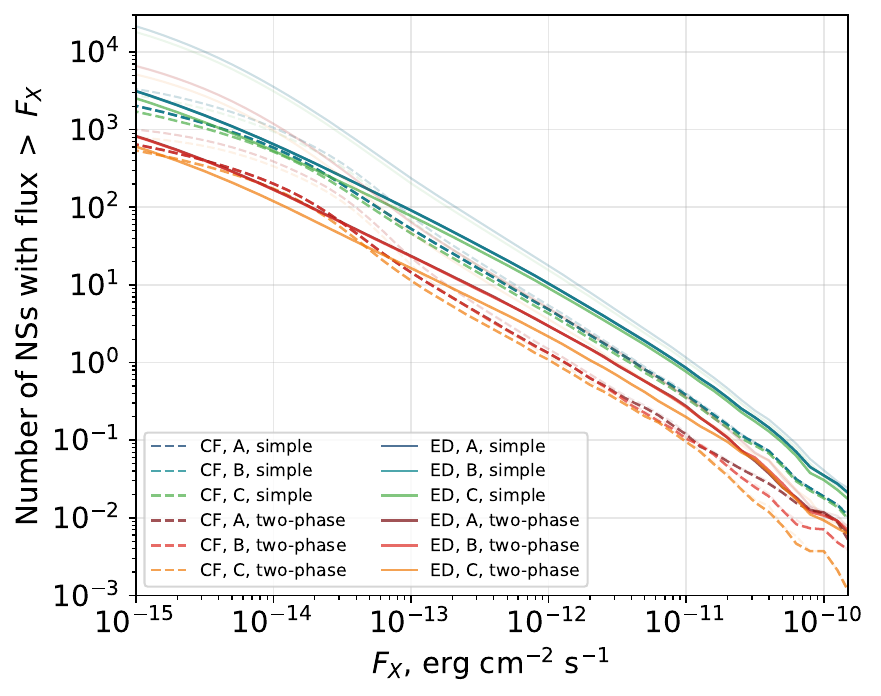}
    \caption{Number of dAINSs in the Milky Way with fluxes exceeding $F_\text{X} = 10^{-15}$~erg~s$^{-1}$~cm$^{-2}$. The total number of INSs in the Galaxy is $3\times10^8$. Faint lines with the same linestyles show this number without interstellar absorption.
    }
    \label{fig:flux}
\end{figure}

The effective temperature of the polar caps $T$ and the X-ray flux $F_\text{X}$ of the dAINSs for the distant observer are shown in Figures~\ref{fig_six} and~\ref{fig:flux}. The temperature depends very weakly on the magnetic field and the accretion rate
\begin{equation}
    T = 4.2\times10^6 \, \left(\frac{{\Bs}}{10^{12}\text{~G}}\right)^{1/7}\left(\frac{\dot{M}}{10^{11}\text{~g~s}^{-1}}\right)^{5/28}\text{~K}.
\end{equation}
Thus, despite the fact that NSs can have accretion discs with a wide range of dipole surface magnetic field values ${\Bs}$, varying between $10^7$ and $10^{13}$~G, and the accretion rates of $10^{8}-10^{12}$~g~s$^{-1}$, the temperature does not vary that much and is expected to be in the range $\sim4\times10^6-10^7$~K. The maximum of the corresponding blackbody spectrum is at $1-2.4$~keV. The results differ slightly in the two magnetic field models. In the model with the exponentially decaying field, the NSs have lower magnetic fields, the Alfv{\'e}n radius is smaller, and the polar caps are larger. Consequently, the same accretion energy is released from a larger area in the ED model than in the CF model, resulting in a lower effective temperature.

Figure~\ref{fig:flux} illustrates the number of NSs with accretion discs $N_\text{X}$ with the X-ray fluxes higher than a given value $F_\text{X}$. 
The number of potentially observable sources -- the objects with $F_\text{X}\gtrsim10^{-15}$~erg~s$^{-1}$~cm$^{-2}$ -- differs between the two ISM models. In the case of the one-phase model, it is $\approx1600-3000$, while in the two-phase model it is $\approx500-1000$. This is because there are more dAINSs in the one-phase model. Also, in the one-phase model, the AINSs can have accretion discs when they are located higher above the Galactic plane, where interstellar absorption is lower. These two effects result in the difference between the two ISM models.

The slope of the lines is influenced by the spatial distribution of dAINSs near the position of the Sun in the Galaxy. If the sources are distributed over a filled sphere around the Sun, their number depends on the radius of the sphere as $N_\text{X}\propto d^3$, and if the sources are distributed in a thin disc, then the dependence on the radius of the disc is $N_\text{X}\propto d^2$. In both cases, the flux of a source decreases with the distance in the same way, $F_\text{X}\propto d^{-2}$. This gives a slope coefficient $\kappa=\log N_\text{X} / \log F_\text{X} = -1.5$ in the former case and $\kappa = -1$ in the latter. In our modeling, the NSs with the fluxes $\sim10^{-11}-10^{-10}$~erg~s$^{-1}$~cm$^{-2}$ are located near the position of the Sun, and the absolute value of their slope coefficient is $|\kappa| \approx 1.3-1.5$, $\kappa<0$. Fainter sources with fluxes $F_\text{X}\sim 10^{-15}-10^{-13}$~erg~s$^{-1}$~cm$^{-2}$ are located further from the Sun, and the corresponding coefficient is $|\kappa|\approx 0.7-1$, meaning that the slope is flatter than that for a population in a disc. This is due to interstellar absorption, which becomes important at the distances corresponding to these values of X-ray flux. 

{The estimated number of observable dAINSs depends on the assumption that the X-ray luminosity of an accreting NS is proportional to the Bondi accretion rate. 
If the actual accretion rate onto the NS surface is a fraction $\xi\le1$ of the Bondi rate, the luminosity of each source decreases by the factor $\xi$. Since the cumulative number of sources approximately scales as $N_\text{X}\propto F_\text{X}^{-\alpha}$ with $\alpha\approx1$ (see Fig.~\ref{fig:flux}), the number of observable sources is then reduced by approximately the same factor $\xi$.}

The luminosity of a dAINS with the accretion rate of $\dot{M}\sim10^{10}-10^{12}$~g~s$^{-1}$ is $L_\text{X}\sim2\times10^{30}-2\times10^{32}$~erg~s$^{-1}$. If its X-ray flux is $F_\text{X}\sim10^{-13}$~erg~s$^{-1}$~cm$^{-2}$, then the distance to the source is $\sim0.35-3.5$~kpc. Here, we calculate the distance using eq.~\ref{eq_f_0} without accounting for interstellar absorption. As the dAINSs are located in the Galactic disc, interstellar absorption plays a role.
Taking interstellar absorption into account, a bright source with an accretion rate of $\dot{M}=10^{12}$~g~s$^{-1}$ at a distance $3.5$~kpc would be visible with $F_\text{X} \gtrsim 5\times10^{-14}$~erg~s$^{-1}$~cm$^{-2}$. At this distance, the sources with accretion rates $\dot{M}\lesssim 3\times10^{10}$~g~s$^{-1}$ are on the verge of visibility with fluxes of $\lesssim10^{-15}$~erg~s$^{-1}$~cm$^{-2}$. Therefore,  the dAINSs could be visible at distances up to several kpc.

The same calculations, when performed for MHD turbulence with $v_\text{t}(r)\propto r^{1/2}$ instead of the Kolmogorov scaling $v_\text{t}(r)\propto r^{1/3}$, yields a reduced number of dAINSs, shown in Table~\ref{tab_MHD}.

\begin{table}
\centering
\caption{The number of accretors with discs in the Galaxy, calculated in different models, assuming the MHD turbulence in the ISM ($\alpha=1/2$). The propeller model D and the model CF (constant field) are not shown because they yield an insufficient number of dAINSs. The ratio of the accreting INSs to the total number of AINSs in the Galaxy ($N_\text{A}/N$, $N=3\times10^8$) remains the same as in Table~\ref{tab:results}.
}
\centering
\label{tab_MHD}
\renewcommand{\arraystretch}{1.1}
\begin{tabularx}{\columnwidth}{@{\extracolsep{\fill}} l p{3.5cm} p{2.9cm}}
\bottomrule
\addlinespace[4pt]
\multicolumn{1}{l}{} & \multicolumn{2}{l}{One-phase ISM, exponentially decaying field} \\
\midrule
Propeller model & $N_\text{D}$ & $N_\text{D}/N_\text{A}$, \% \\
A & $(3.15 \pm 0.26)\times10^5$ & $0.235 \pm 0.021$\\
B & $(1.6 \pm 0.6)\times10^4$ & $0.0160 \pm 0.0016$\\
C & $(2.64 \pm 0.24)\times10^5$ & $5.5 \pm 0.3$\\
\bottomrule
\addlinespace[4pt]
\multicolumn{1}{l}{} & \multicolumn{2}{l}{Two-phase ISM, exponentially decaying field} \\
\midrule
Propeller model & $N_\text{D}$ & $N_\text{D}/N_\text{A}$, \% \\
A & $(8.0 \pm 1.0)\times10^4$ & $0.097 \pm 0.011$\\
B & $(4.2\pm 2.1)\times10^3$ & $0.0260 \pm 0.0023$\\
C & $(6.5 \pm 1.1)\times10^4$ & $0.060 \pm 0.006$\\
\bottomrule
\end{tabularx}
\end{table}

The ratio of the turbulent to Keplerian specific torques in this case is
\begin{equation}
    n_\text{d}'\approx 0.43~ \left(\frac{v}{10\text{~\kms}}\right)^{-24/7} \left(\frac{B}{10^{12}\text{~G}}\right)^{-2/7} \left(\frac{n}{1\text{~cm}^{-3}}\right)^{1/7}.
\end{equation}
This value is approximately one order of magnitude lower than $n_\text{d}$ for the Kolmogorov spectrum (eq.~\ref{n_kol}), implying that it is harder for AINSs to acquire an accretion disc.

In the ED model, the number of dAINS is now 1-2 orders of magnitude smaller, while in the CF model, dAINS are completely absent. The maximum values of the magnetic field and the characteristic velocity of dAINSs are also changed. The parameters now are ${\Bs}\lesssim10^{11}$~G and $v\lesssim20$~\kms, whereas in the model with a Kolmogorov turbulence spectrum the boundaries are far less restrictive: $v\lesssim40-50$~\kms and ${\Bs}\lesssim10^{13}-10^{14}$~G. Because the magnetic field for the dAINSs is $\sim10^{11}$~G, no accreting INSs acquire discs in the model with the constant field; it can develop a disc only if its magnetic field has decayed. 

The typical parameters change with the turbulence model. For the NSs with exponentially decaying magnetic fields, the ${\Bs}$ distribution is narrower than in Figure~\ref{fig_six}, spanning ${\Bs}\sim10^7-10^{11}$~G, so possible effective temperature values shift to $T\sim6\times10^5-5\times10^6$~K. The narrower velocity distribution leads to a narrower $|z|$-distribution, both are now similar to the $v$ and $|z|$-distribution in the CF model with Kolmogorov turbulence, shown in Figure~\ref{fig_six}.

\section{Application to long-period transients}
\label{sec:lprt}

LPTs are a recently discovered class of pulsating radio sources with periods of $\sim10^2-10^5$~s \citep{2022Natur.601..526H, 2023Natur.619..487H}. 
Up to now, about 15 sources are known \cite[see][for a review]{2026JHEAp..5200566R}. 
In several cases, they have been identified as close binary systems consisting of a white dwarf and a low-mass companion (an M-dwarf) \citep{2024ApJ...976L..21H, 2025NatAs...9..672D, 2025A&A...699A.341B, 2025MNRAS.542.1208A}. 
For several other sources, it is suspected that they also belong to this class \citep{2025MNRAS.542..203M, 2025PASA...42..129H}. However, for roughly half of the known LPTs, their characteristics are consistent with isolated NSs \citep{2023AstL...49..553A, 2024ApJ...961..214R}, in which case the observed periods likely represent the stellar spin period. 
{ The emission mechanism operating in this framework is not known, and various models are discussed in the literature; see, e.g., \cite{2023MNRAS.520.1590S,2024MNRAS.533.2133C, 2026arXiv260808243Q} and references therein.}
Long spin periods may be reached either due to propeller interaction with a fallback disc \citep{2024ApJ...967...24F} or magnetospheric twist injections \citep{2026ApJ..1000...55S}.
 
Recently, \cite{ferrario25} proposed an alternative scenario. 
In this framework, some subset of emitting objects may be AINSs, and radiation is produced by the electron cyclotron maser emission mechanism (ECME). 
In this section, we develop the approach by \cite{ferrario25} and discuss the possibility that a fraction of LPTs can be related to dAINSs. 
{We emphasise, however, that we are not necessarily advocating for this emission mechanism or claiming that most LPTs are isolated NSs. 
Indeed, there is growing evidence that white dwarfs seem to be the dominant class, where ECME mechanisms can also operate \citep{2026arXiv260808243Q}.
It is worth exploring alternatives involving neutron stars, however, as several LPTs show no evidence for a binary nature, as we discuss below.
}

\begin{figure}
	\includegraphics[width=\columnwidth]{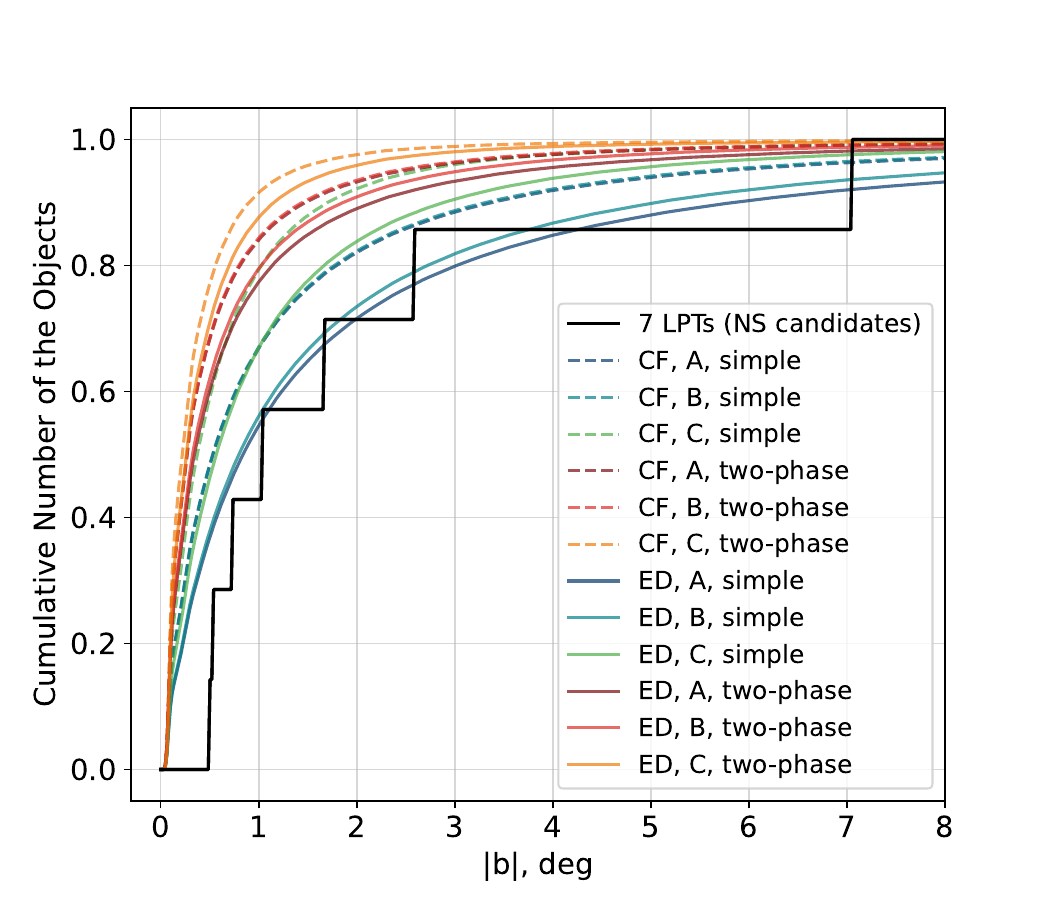}
    \caption{Normalised cumulative density distribution of several observed LPTs with NS candidates over galactic latitude and the distribution of the accreting NSs with discs calculated within 12 models. 
    }
    \label{fig:lpt}
\end{figure}

At first, we compare the $|z|$-distribution of dAINS with those LPTs that are not obviously related to white dwarfs in binary systems. 
The spatial distributions of observed LPTs and calculated isolated accreting NSs with discs are shown in Figure~\ref{fig:lpt}. The plotted LPTs include seven sources: GLEAM-X J1627-5235, ASKAP J1935+2148, CHIME J0630+25, ASKAP J1839-0756, ASKAP 1424-6126, ASKAP 1651-4505, and ASKAP 1700-4457. The values of their Galactic latitudes are taken from Table~1 of the review by \cite{2026JHEAp..5200566R} and from \cite{2026arXiv260307857P,2026arXiv260620067W} for newly discovered sources.
For the calculated distribution of the dAINSs, we present results for two magnetic field models (constant field CF and exponential decay ED), three propeller models (A, B, and C), and two models of the ISM distribution (simple and two-phase). The propeller model D yields a negligible number of AINSs and is therefore not shown.

In Figure~\ref{fig:lpt}, we see that the majority of dAINSs are expected to be situated close to the Galactic plane. For models with the exponential field decay, propeller models A and B, and one-phase ISM, the curves closely follow the observed distribution (note, however, the poor statistics of the LPTs). 

 The farthest LPTs in our observational sample are situated (accounting for uncertainties) at distances $\lesssim 4-5$~kpc from the Sun \citep{2026JHEAp..5200566R}, and our modeling suggests that the number of dAINS in this region is $\sim 10^4 - 3\times 10^5$. The duty cycles of the observed sources are $\lesssim$~a few percent. 
 Thus, the probability of detecting a given source accounting for the beaming is $\gtrsim 0.01$.
Altogether, the expected number of potentially detectable 
LPTs related to dAINSs is $\sim 100 - 10^4$. Some of them can be dim radio sources (see below), and so they avoid detection. Some can be transient because of the disc disappearance (e.g., due to fluctuations of the ISM density or properties of the turbulence). Here, we just demonstrate that the number of potential sources is high enough to explain the number of already discovered sources. 

As discussed in Sec.~\ref{sec:acc}, INSs start to accrete at periods $P_\text{A}\gtrsim 10^3$~s, see Eq.~(\ref{eq:pa}). This value is in rough correspondence with the shortest periods of LPTs \citep{2026JHEAp..5200566R}. Then, the AINS spins down and can reach very long periods, as mentioned by \cite{ferrario25} following arguments by \cite{2002A&A...381.1000P}. However, for a reasonable amount of time ($\sim10^7$~yr; see Sec.~\ref{sec:acc}), the spin period is $\lesssim10^4$~s. Thus, we can expect that $\sim 0.1-1$\% of dAINSs have spin periods within a few hours, i.e., in the range measured for LPTs.  

Note that due to the transient behavior of the disc, we do not expect that the usual 
equilibrium period for the disc accretion, $p_\text{eq}\approx 1000 \, L_{30}^{3/7} \mu_{30}^{6/7}\, \text{s}$, can be reached. The characteristic time related to turbulence ($\sim R_\text{G}/v$), which is from a few months up to a few years, is much shorter than the time necessary to reach the equilibrium $\sim p/\dot p= (2\pi I)/(\dot M\sqrt{GMR_\text{A}} p) \sim 10^5-10^7 $~yr. 

\subsection{Cyclotron-maser instabilities in dAINSs} \label{sec:maserinstabs}

We here briefly recap the mechanism detailed by \cite{ferrario25} for ECME activation for compact stars surrounded by accretion discs \cite[see also][]{2025ApJ...981...34Q,2026ApJ...999L...2Z}.

As the electrons (or positrons) flow towards the star from the disc, the field intensity increases and the particles acquire perpendicular velocity (relative to $\boldsymbol{B}$; i.e., cyclotronic spiraling).
This causes the parallel velocity to decrease if the first adiabatic invariant is conserved, until eventually the waves emitted from the accelerating charges reach resonance with the gyrofrequency. 
The radius at which this happens depends on the initial ``pitch angle'', $\alpha$, between the velocity components: an electron with an angle smaller than a critical value -- defined such that the particle would hit the surface before resonance --  falls into the so-called loss cone. 
At the resonance radius, if not in the cone, the mirror force, $F_{||} = \mu_e dB/ds$ where $\mu_e$ is the particle's magnetic moment and $dB/ds$ is the change of magnetic field strength along the path, ejects them outwards.
Depending on the distribution of velocities for particles injected from the disc, particles may clump at certain radii such that instabilities trigger the ECME.
The frequency of the radiation is set only by the local field strength if the ratio of the plasma frequency to gyrofrequency is small (see equation A11 therein). 
Note that this requires a small particle number density $n_e$ to not scatter (explaining why such instabilities do not occur in typical low-mass X-ray binaries, for instance).

Maximum wave growth requires the electric field to rotate at the same frequency and direction as the emitting electrons, producing light that is circularly polarised. 
Through Faraday rotation within the surrounding medium, these waves may be subsequently converted into linear polarisation \citep{ferrario25}.
In general, the phase shift scales as $\nu^{-3}$, which implies linear polarisation at low frequencies and circular at high frequencies. 
Loss-cone particles hit the surface, maybe making a small hotspot (like in ASKAP J1832). 
Overall, such a model matches observations of LPTs rather well \cite[see Table 2 in][]{ferrario25}.

In what follows, we explore how conditions set by dAINSs may be conducive to the ECME or otherwise.

\subsection{Quenching conditions for cyclotron maser instabilities} \label{sec:quenching}

Consider an axisymmetric flux tube connecting the surface of an accreting NS to the outer edge of the associated disc.
Since the magnetic flux, $\Phi = \boldsymbol{B} \cdot \boldsymbol{A}$, should be approximately conserved along a bundle of field lines we can write
\begin{equation}
B(r_1, \theta_1) A(r_1, \theta_1) = B(r_2, \theta_2) A(r_2, \theta_2),
\end{equation}
for any two points (labeled 1 and 2) along the same magnetic flux tube.
As the magnitude of the dipole field at any point reads $B(r,\theta) = {\Bs R^3}/r^3 \sqrt{1 + 3 \cos^2\theta}$, we have
\begin{equation} \label{eq:arealaw}
    A_2 = A_1 \left( \frac{r_2}{r_1} \right)^3 \sqrt{\frac{1 + 3\cos^2\theta_1}{1 + 3\cos^2\theta_2}}.
\end{equation}
If plasma is injected from an equatorial disc ($\theta_2 = \pi/2$) at the Alfv{\'e}n radius \eqref{eq:alfvenradius}, we can use expression \eqref{eq:arealaw} to find the cross-sectional area associated with any given radius
\begin{equation} \label{eq:tubearea}
A(r) = \frac{2 \pi r^3 h}{\RA \sqrt{4 - 3 r/\RA}},
\end{equation}
where $h = H/r$ is a dimensionless parameter quantifying the thickness of the tube; for a thick disc we expect $h \sim 1$. 
Later on we will associate $r_1$ with the altitude of the maser emission site ($\rnu$).
In deriving the above we made use of the relation $A_{2} =  2\pi h \RA^2$ (for the \emph{cross-sectional} area) and the fact that the radius and latitude at any point along that path satisfies $r = \RA \sin^2\theta$ (through the dipolar assumption) to eliminate angles.

To progress, we assume that the accreting plasma is fully ionized and macroscopically neutral. 
That is we assume $n_\text{e} = n_\text{p}$, implying the mass-density within the tube is just
\begin{equation}
\rho = m_\text{e} n_\text{e} + m_\text{p} n_\text{p} \approx n_\text{p} m_\text{p} = n_\text{e} m_\text{p}.
\end{equation}
As such, the continuity equation allows us to relate the mass accretion rate $\dot{M}$ to the local electron density $n_\text{e}$ and the free-fall velocity $v_{\text{ff}}$ through \citep{do73}
\begin{equation}
    \dot{M} = \rho v_{\text{ff}}(r) A(r) = m_\text{p} n_\text{e}(r) v_{\text{ff}}(r) A(r) ,
\end{equation}
Thus, the local number density at any given point is
\begin{equation}
\label{eq:ne_base}
    n_\text{e}(r) = \frac{\dot{M} \sqrt{r}}{m_\text{p} \sqrt{{2GM}} A(r)}.
\end{equation}

\begin{figure}
	\includegraphics[width=\columnwidth]{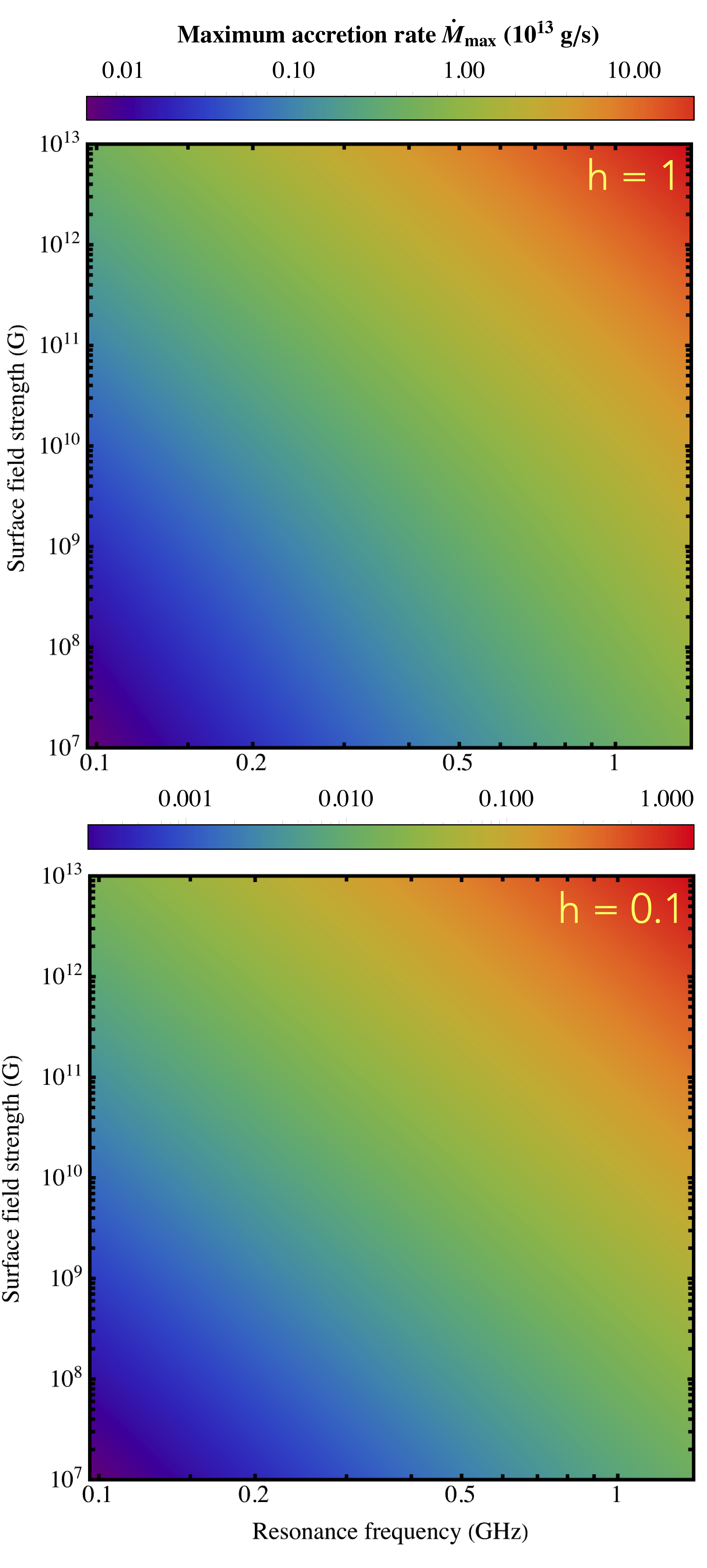}
    \caption{Maximum accretion rate from expression \eqref{eq:finalquench} for canonical choices $\delta^2  = s = 1$ with thickness $h=1$ (top panel) or $h=0.1$ (bottom) as a function of frequency and surface field strength; redder shades indicate greater $\dot{M}_{\rm max}$.}
    \label{fig:quench}
\end{figure}

Following \cite{ferrario25}, we impose that ECME is quenched when the local plasma frequency, $\omega_p = {{4\pi n_\text{e} e^2}/{m_\text{e}}}$, exceeds a fraction $\delta$ of the cyclotron frequency, $\Omega_e=eB(r,\theta)/m_\text{e} c$, i.e., when\footnote{We assume Newtonian flows here to avoid carrying around factors of the Lorentz factor, $\gamma \approx 1$, as this can be effectively absorbed into the parameter $\delta$.}
\begin{equation} \label{eq:plasmaquench}
    \omega_p > \delta  \Omega_e.
\end{equation}
The dimensionless scaling factor $\delta$ is typically taken to lie in the range $0.1 \lesssim \delta \lesssim 1$ \cite[see, e.g.,][]{md82}.
By substituting the dipole field expression into $\Omega_e$ we find
\begin{equation}
    2\pi \nu = \frac{s e \Bs R^3}{m_\text{e} c \rnu^3} \sqrt{1 + 3\cos^2\theta}.
\end{equation}
For emission constrained to the accretion funnel anchored at the Alfv{\'e}n radius, the local coordinates are again related by $\cos^2\theta = 1 - r/R_A$ and we then can get a self-consistent but transcendental equation for the emission radius 
\begin{equation}
\rnu = R \left[ \frac{s e \Bs \sqrt{4 - 3\rnu/\RA}}{2\pi m_\text{e} c \nu} \right]^{1/3}.
\label{eq:r_nu}
\end{equation}
Note that we generally require $R <\rnu < \RA$ else particles fall into the loss cone.
In the above, we have introduced the symbol $s$, as the emission is associated with the $s$-th harmonic of the cyclotron frequency (that which Ferrario calls $n$ and we call $s$ to avoid confusion). 
We may now substitute the definitions for $\omega_p$ and $\Omega_e$ into condition \eqref{eq:plasmaquench} to find
\begin{equation} \label{eq:ncrit}
    n_{\text{crit}} = \frac{\pi m_\text{e} \delta^2 \nu^2}{s^2 e^2},
\end{equation}
for the critical density $n_{\text{crit}}$ above which the maser is quenched.
Equating this critical value \eqref{eq:ncrit} with that from expression \eqref{eq:ne_base} at $r = \rnu$ we get
\begin{equation}
    \dot{M}_{\text{max}} = \frac{\pi m_\text{e} m_\text{p} \delta^2 \nu^2 \sqrt{2GM}}{s^2 e^2 \sqrt{\rnu}} A(\rnu),
\end{equation}
for the quenching accretion rate.
Finally, substituting the definition for the Alfv{\'e}n radius \eqref{eq:alfvenradius} and the cross-sectional area \eqref{eq:tubearea} we arrive at an implicit quenching condition for $\dot{M}$ in the form
\begin{equation} \label{eq:finalquench}
    \dot{M}_{\rm max} \sqrt{4 - \frac{3\rnu}{\RA}} = \frac{2\pi^2  m_\text{e} m_\text{p} \sqrt{2GM} h \delta^2 \nu^2 \rnu^{5/2}}{s^2 e^2 \RA}.
\end{equation}
Equation \eqref{eq:finalquench} can be checked for any given combination of parameters from the population synthesis. 
For example, if we set $\nu = 1$~GHz, $\Bs = 10^{12}$~G, $ \delta =1$, $h=1,$ and $s = 1$ for simplicity along with canonical neutron-star macroscopics ($R = 10^{6}$~cm, $M = 1.4 M_{\odot}$), we find $\dot{M}_{\rm max} \approx 10^{14} \text{ g s}^{-1}$. 
By contrast, taking $\delta = 0.3$ gives $\dot{M}_{\rm max} \approx 3 \times 10^{12} \text{ g s}^{-1}$. 
For $B = 10^{9}$~G we get instead the tighter requirement $\dot{M}_{\rm max} \sim 10^{11} \text{ g s}^{-1}$. 
This is physically intuitive: a larger field implies a larger magnetospheric volume and thus more particles are required to choke any local region. 
More generally, the relationship \eqref{eq:finalquench} is depicted in Figure~\ref{fig:quench} for the range of dAINS parameters depicted in Fig.~\ref{fig_six}.

We see that, for the peak of the distribution shown in Fig.~\ref{fig_six} with $\Bs \sim 10^9$~G, we expect a maximum accretion rate of $\approx 10^{11} \text{g s}^{-1}$ for a thick disc with $h=1$ or a smaller value of $\approx 10^{10} \text{g s}^{-1}$ if $h = 0.1$ at an emission frequency of $\approx 140$~MHz. 
The latter value, in particular, resides to the left of the peak of the accretion rate obtained from the population synthesis (middle top panel of Fig.~\ref{fig_six}) and thus we expect only a modest fraction of dAINS to manifest as LPTs if the disc is not overly thick, though such a relationship depends crucially on the correlation between $\dot{M}$ and $B$ values obtained from the synthesis. 

\begin{figure}
	\includegraphics[width=0.98\columnwidth]{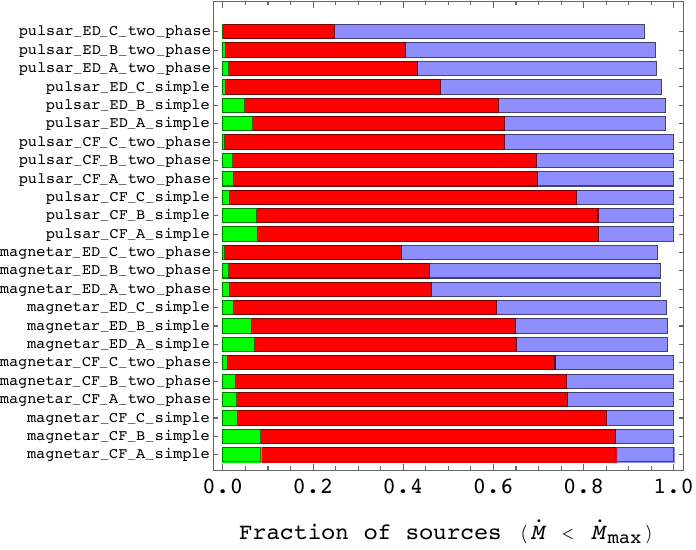}
    \caption{Fraction of accreting dAINSs such that the accretion rate lies below the quenching maximum \eqref{eq:finalquench}  from expression \eqref{eq:fraction}.
    Cases in blue correspond to $h \delta^2=1$, red to $h \delta^2=0.1$, and green to $h \delta^2=0.01$.}
    \label{fig:fractions}
\end{figure}

Such a check can be made directly by taking the raw data and fitting joint probability distribution functions (PDFs) for each model, $\mathcal{P}_{\rm model}(B,\dot{M})$, normalised such that $\iint dB d\dot{M} \mathcal{P}_{\rm model} = 1$. 
Introducing an indicator function -- defined via $\boldsymbol{1}_{A}(\boldsymbol{x}) = 1$ if $\boldsymbol{x} \in A$ and 0 otherwise -- we can investigate the fraction satisfying the quenching condition \eqref{eq:finalquench} by computing
\begin{equation} \label{eq:fraction}
    f(\text{Model}) = \iint dB d\dot{M} \mathcal{P}_{\rm model}(B,\dot{M}) \times \boldsymbol{1}_{\dot{M}<\dot{M}_{\rm max}(B)}.
\end{equation}
The results are shown in Figure~\ref{fig:fractions} for various combinations of the effective thickness $h \delta^2$. For realistic cases with thick discs such that $h \delta^2 \lesssim 0.1$, we see that roughly half of the sources would be expected to be able to trigger ECME for any model ($f \lesssim 0.5$).
For thin discs, the results are more pessimistic: for exponentially-decaying pulsar fields, the instability may fire only in $f \leq 1\%$ of cases. 
For the decaying, two-phase pulsar model C, we see in fact that almost no sources can satisfy the condition if $h \delta^2 = 10^{-2}$.
In general, however, stronger fields permit a larger Alfv{\'e}n radius and thus a larger volume for the ECME to operate, yielding a systematically larger $f$ for magnetar models.
While the model presented here cannot self-consistently account for disc thickness or predict it, this would be an avenue worth exploring in future to make more refined predictions in this respect.

Note that for frequencies below the plasma frequency, travelling waves become evanescent and radio emissions are exponentially suppressed \cite[see, e.g., Sec. 3.2.1 of][for a general discussion]{2020PTEP.2020j3E01W}.
In a disc model, there exist regions that are still vacuum-like (near the poles) and radiation can escape freely, though for spherical accretion all regions are choked to some degree and even a very low $\dot{M}$ can prevent radio wave escape.


\subsection{Radio spectrum} \label{sec:spectra}


Aside from quantifying quenching, predictions for radio spectra can also be made in the model.
Suppose that non-thermal particles are injected into the magnetosphere at the Alfv{\'e}n radius with a pitch angle $\alpha_\mathrm{A}$ relative to the local magnetic field. 
For an isotropic injection, the PDF takes the classical form
\begin{equation} \label{eq:pitch}
    \mathcal{P}(\alpha_\mathrm{A}) = \cos\alpha_\mathrm{A}.
\end{equation}
More generally, we can keep the function $\mathcal{P}$ free to investigate how spectra may vary depending on the properties of the disc. 
In particular, we expect $h \sim 1$ following arguments made earlier; this will affect injection probabilities in ways we discuss below.

Adopting expression \eqref{eq:pitch} for the moment though, we have, by conservation of the first adiabatic invariant, that $\sin^2\alpha / B$ is conserved \citep{ferrario25} and a particle mirrors when its pitch angle reaches $\alpha = \pi/2$ by definition. 
Therefore, the local magnetic field, $B_\mathrm{m}$, at the mirroring radius, $R_{||}$, reads
\begin{equation}
 B_\mathrm{m} = \frac{B_\mathrm{A}}{\sin^2\alpha_\mathrm{A}},
\end{equation}
where we note that $\left({R_\mathrm{A}}/{R_{||}}\right)^3 = 1/{\sin^2\alpha_\mathrm{A}}$ for a dipolar field, and hence $R_\mathrm{m} = R_\mathrm{A} (\sin\alpha_\mathrm{A})^{2/3}$.

To find the spatial distribution of the mirror shells, we need to calculate the cumulative probability that a particle mirrors at a radius smaller than $r$. 
This corresponds to particles injected with pitch angles smaller than $\alpha(r)$, i.e.,
\begin{equation}
    \mathbb{P}(R_{||} < r) = \int_{0}^{\alpha(r)} \cos\alpha_\mathrm{A} \, d\alpha_\mathrm{A} = \sin\alpha(r),
\end{equation}
where we have used expression \eqref{eq:pitch}.
Since $\sin\alpha(r) = (r/R_\mathrm{A})^{3/2}$, the PDF for the location of the mirror points is simply
\begin{equation}
    \frac{d\mathcal{P}}{dR_{||}} =  \frac{d}{d R_{||}} \left[ \left( \frac{R_{||}}{R_\mathrm{A}} \right)^{3/2} \right] = \frac{3}{2 R_\mathrm{A}} \left( \frac{R_{||}}{R_\mathrm{A}} \right)^{1/2}.
\end{equation}
Since $d\mathcal{P}/dR_{||} \propto \sqrt{R_{||}}$, the majority of injected particles mirror in the outer magnetosphere near $R_\mathrm{A}$, leading to a starved inner magnetosphere.
This naturally explains why there is no X- or gamma-ray emissions from the ECME even for highly-magnetised neutron stars: there are simply too few particles mirroring in regions where the field is strongest.

For an arbitrary pitch distribution instead of expression \eqref{eq:pitch}, the steps can be easily repeated, and the above logic carries through. 
Moreover, it is instructive to instead consider the spectrum of emissions rather than the probabilities associated with mirror points.
Denoting the cyclotron frequency at $R_\mathrm{A}$ through $\nu_\mathrm{A} = {e B_\mathrm{A}}/{2\pi m_e c}$, we invert the dipole relation again to get the injection pitch angle corresponding to a given emission frequency,
\begin{equation}
    \alpha_A(\nu) = \arcsin\left( \sqrt{\frac{\nu_\mathrm{A}}{\nu}} \right),
\end{equation}
from which we deduce the fraction of radiation emitted in a band $d\nu$ through
\begin{equation}
    S(\nu) \propto \left| \frac{d\mathcal{P}}{d\nu} \right| = \mathcal{P}(\alpha_\mathrm{A}) \left| \frac{d\alpha_\mathrm{A}}{d\nu} \right|.
\end{equation}
As such, the generalised spectral flux distribution for any injection PDF is
\begin{equation}
    S(\nu) \propto \mathcal{P}\left( \arcsin \sqrt{ \frac{\nu_\mathrm{A}}{\nu} } \right) \frac{\sqrt{\nu_\mathrm{A}}}{2\nu\sqrt{\nu - \nu_\mathrm{A}}}.
\end{equation}
Suppose we adopt a parametrisation of the form 
\begin{equation} \label{eq:generalpitch}
    P(\alpha_\mathrm{A}) \propto \sin^\ell \alpha_\mathrm{A} \cos^m \alpha_\mathrm{A},
\end{equation} for some constants $\ell$ and $m$. 
Using $\sin \alpha_\mathrm{A} = \sqrt{\nu_\mathrm{A} / \nu}$ and $\cos \alpha_\mathrm{A} = \sqrt{(\nu - \nu_\mathrm{A}) / \nu}$, the predicted spectrum is thus
\begin{align}
    S(\nu) &\propto \left( \sqrt{\frac{\nu_\mathrm{A}}{\nu}} \right)^\ell \left( \sqrt{\frac{\nu - \nu_\mathrm{A}}{\nu}} \right)^m \frac{\sqrt{\nu_\mathrm{A}}}{2\nu\sqrt{\nu - \nu_A}} \nonumber \\
    &\propto \nu^{-\frac{\ell+m+2}{2}} (\nu - \nu_\mathrm{A})^{\frac{m-1}{2}}.
\end{align}

For high frequencies above the truncation value ($\nu \gg \nu_\mathrm{A}$), we approximate $(\nu - \nu_\mathrm{A})^{\frac{m-1}{2}} \approx \nu^{\frac{m-1}{2}}$ and thus
\begin{equation} \label{eq:predictionhighfreq}
    S(\nu \gg \nu_\mathrm{A}) \propto \nu^{-\frac{\ell+3}{2}},
\end{equation}
for any $m$.
It could be argued that large $\ell$ is the natural choice for neutron stars surrounded by thick discs that are accreting somewhat moderately. 
In such a case, high-energy particle injection is likely to be driven by magnetic reconnection within equatorial current sheets and acceleration across quasi-perpendicular shocks. 
Both of these mechanisms generate strong electric fields that accelerate particles predominantly across the local field lines, naturally skewing the injected momentum to be highly perpendicular ($v_\perp \gg v_\parallel$). 
As a result, electrons are therefore injected with pitch angles that are concentrated towards $\alpha_\mathrm{A} \approx \pi/2$; this corresponds to a large value of $\ell$. 

\begin{figure}
	\includegraphics[width=\columnwidth]{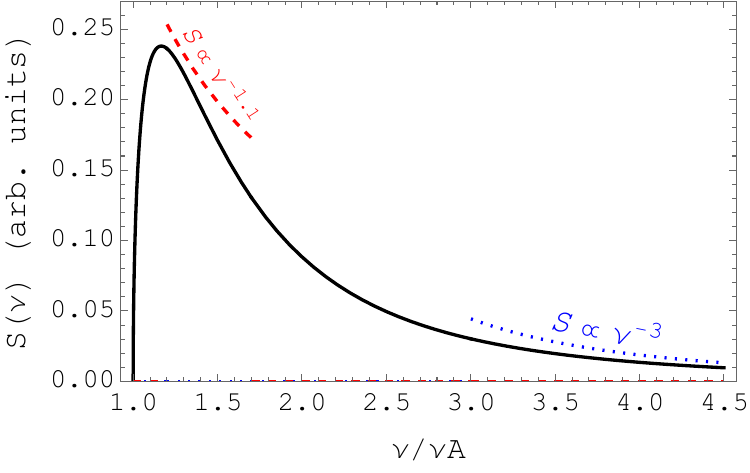}
    \caption{Normalised frequency spectrum, $S(\nu)$, for the ECME operating in a neutron-star-plus-disc system with pitch angle injections following expression \eqref{eq:generalpitch} with $m=2$ and $\ell=3$; overlaid are power-law scalings.}
    \label{fig:spec}
\end{figure}

For GLEAM-X J1627, \cite{2022MNRAS.514L..41E} finds a best-fitting power-law index of $\approx -3$ at emission frequencies above $\approx 200$~MHz.
Such an index matches the prediction \eqref{eq:predictionhighfreq} if $\ell \gtrsim 3$, meaning that the model can naturally explain the emission spectrum of at least some LPTs.
The global spectrum for $m=2$ and $\ell=3$ is shown in Figure~\ref{fig:spec}. 
For these particular choices, we have the relations $S(\nu) \propto \nu^{-1.1}$ at frequencies just above the cutoff at $\nu_A$, again roughly consistent with the spectrum from GLEAM-X J1627 \cite[see Figure 1 in][]{2022MNRAS.514L..41E}.

Although detailed modeling would be required beyond the scope of this paper to connect disc characteristics to the spectrum, these considerations highlight that not only can the population density be roughly reproduced with a dAINS model but that spectra can be recovered with reasonable assumptions.

\section{Discussion}
\label{sec:discuss}

%

\subsection{Uncertainties in the population synthesis scenario and disc properties}

 Typically, population synthesis models contain many simplifications, which are necessary to perform calculations for many objects with various initial parameters and evolutionary paths. The model presented above is not an exception. Uncertainties related to the population synthesis of isolated NSs have been discussed by \cite{2026JHEAp..5300643A}. Here, we briefly summarise key uncertainties related to discs around AINSs. 

  The first group of issues concerns disc formation. The angular momentum captured from the ISM is quite low. Its value depends on the properties of interstellar turbulence. The latter are quite uncertain on the scale of the Bondi radius. If we overestimate the available angular momentum even in our pessimistic scenario, then the number of discs might be lower. In addition, the captured angular momentum depends on the size of the Bondi radius, which, in its turn, depends on the spatial velocity of the NS. The velocity (related mainly to the natal kick) is quite uncertain (see a review in \citet{2025NewAR.10101734P}). This one of the main uncertainties of our model.  

  Disc formation also depends on the value of the NS magnetic field. We applied a simple model of the field decay. Up to now, there are no direct data or even strong restrictions on the field evolution of isolated NSs on $\gtrsim$~Gyr time scales. If the field does not decay significantly, the number of discs is lower, too. To avoid this uncertainty, we also considered the scenario with constant magnetic fields. It resulted in the number of dAINSs lower by a factor $\lesssim 10$. 
  However, in this case, the result is sensitive to the initial magnetic field distribution, which is not well-known, especially for lower fields. 

  As the angular momentum of the captured matter is relatively small, significant cooling is necessary for the formation of a thin disc. Our estimates (see Sec.~\ref{sec:acc}) demonstrate that the cooling time scale is comparable with the scale of variation of the external angular momentum ($\sim R_{\mathrm G}/V$). Thus, our expectation is that the disc is not thin ($h\sim 1$). However, detailed calculations of the disc properties require direct modeling, which is beyond the scope of the present study.  
  The same can be said about the uncertainties related to a low accretion rate (and so, low disc mass and low density in the disc). Properties of such discs are not well-known. 

  The question of low accretion rate the second group of questions. They are related to the accretion rate onto the surface of the NS. In the first place, the accretion rate can be inhibited by the magnetic field \citep{2012MNRAS.420..810T}. In addition, we expect the accretion flow to be in the settling-accretion regime \citep{2012MNRAS.420..216S}. This might also influence the disc properties and appearance.

  Regarding the observational appearance of dAINSs, we use very simple assumptions about their spectral characteristics. We applied the assumption of black-body emission from the polar caps of the accreting NS, neglecting any disc contribution.
  This can influence our estimates of the number of observable sources.

  As the discs are expected to be transient, their orientation relative to the NS spin and magnetic axes might change on the scale $\sim R_\mathrm{G}/v$. This would influence the spin evolution of the NS and, probably, the disc properties and appearance. In our modeling, we neglect it, assuming that this is a second-order effect in comparison with general parameters of the discs.

  Finally, regarding the hypothesis that dAINSs can explain part of LPT, we have to note that, as most AINSs are expected to have long spin periods, the model predicts a significant number of transient radio sources without detectable periodicity. In this case, the time scale of the transient activity is $\sim R_\mathrm{G}/v$. 

  { Among uncertainties related to the population synthesis, we mention some weak dependence of our results on the assumed magnetic field distribution and on the fraction of magnetars. These issues have recently been addressed in many studies, e.g. \cite{2019MNRAS.487.1426B, 2025ApJ...986...88S, 2026ApJ...999..102H, 2026NatAs.tmp..204P, 2026arXiv260618608S}. In particular, \cite{2019MNRAS.487.1426B} advocated a much larger fraction of magnetars among newborn NSs, while \cite{2026arXiv260618608S} suggest that magnetars can be described together with other types of NSs with the same (log-uniform) magnetic field distribution. Note that, in the framework of our model, the initial magnetic field distribution is interlinked with the rate of the magnetic field decay. Thus, without a robust description of the magnetic field properties and evolution, we think that it is premature to present results for each of the proposed models. As disc formation around AINSs is favorable for smaller magnetic fields (see Fig.~\ref{fig_six}), our results are more sensitive to the long-term field evolution (for which we apply a simplified approach that is often used in population modeling), not to the initial magnetic field distribution of NSs. Additionally, there is a similar fraction of dAINS that can potentially be LPTs in both magnetar and pulsar populations modeled, as can be seen in Figure~\ref{fig:fractions}. Therefore, the field evolution plays a much bigger role in the number of potential dAINSs and LPTs than the magnetar fraction itself.
  }

\subsection{dAINSs radio luminosity}

In Sec.~\ref{sec:lprt}, we proposed the hypothesis that some of LPTs can be dAINSs. We derived some expected parameters of such sources by developing the approach initiated by \cite{ferrario25}. However, we did not discuss the expected radio luminosity of LPTs.

Radio luminosities of LPTs are not certain for two reasons. One is related to uncertainties in distances to these sources. The second is related to the unknown beaming of radio emission. The beaming can be estimated from the duty cycle of sources. By order of magnitude, it can be estimated as $0.01$. Maximum isotropic radio luminosities derived from the flux measurements and distance are $\sim10^{27}-10^{32}$~erg~s$^{-1}$ \citep{2026JHEAp..5200566R}. Thus, with the beaming $\sim 0.01$ we have to explain luminosities $\sim10^{25}-10^{30}$~erg~s$^{-1}$ 
\cite{ferrario25} only suggested that it can be a fraction of the total energy budget of an AINS: $\sim~\dot M GM/R$, where $R$ is the NS radius. Below, we provide somewhat more elaborate estimates. 

Given that dAINSs accrete low-magnetised cold plasma, the major source of energy for their emission is the potential energy of the accretion flow. Thus, maximal isotropic radio luminosity can be simply estimated as
\begin{equation}
    L_\mathrm{iso,max} \approx \dfrac{GM {\dot M}}{r_\nu},
\end{equation}
%
Assuming simply $r_\nu = 0.1R_\mathrm{A}$ (as $r_\nu < R_\mathrm{A}$), one gets
\begin{equation}
    L_\mathrm{iso,max} \approx 1.2\times 10^{28} \mu_{30}^{-4/7} {\dot M}_{11}\, \mbox{erg s}^{-1}.
    \label{eq:L_radio_max}
\end{equation}
Here $\dot M_{11} = \dot M/(10^{11}$ g s$^{-1} )$.
Hence, for $\sim 1\%$ emission beaming factor, accretion potential energy is capable of supporting luminosity $\sim 10^{30}$ erg s$^{-1}$.

In a more accurate consideration, however, the luminosity estimate could be based on the 
properties of the cyclotron emission mechanism, where local magnetic field $B(r_\nu)$ plays a crucial role as an emission ``catalyzer''. For non-relativistic electrons, the radiated power density (erg s$^{-1}$ cm$^{-3}$) of cyclotron emission is
\begin{equation}
    P_\mathrm{cyc}(r_\nu) \sim \dfrac{1}{4\pi}\, n_e(r_\nu)\, \sigma_\mathrm{T}\, c\, B^2(r_\nu),
\end{equation}
where $\sigma_\mathrm{T}$ is the Thomson cross-section, and $n_e$ the electron number density in the emission region is given by the equation (\ref{eq:ne_base}). Then the isotropic luminosity of the emission region of the volume $\delta V \approx A(r_\nu) H$ reads as
\begin{equation}
    L_\mathrm{ECME} \approx \eta P_\mathrm{cyc}(r_\nu)\, \delta V(r_\nu) = \dfrac{\eta h}{4\pi} \dfrac{\sigma_\mathrm{T}c}{m_\mathrm{p}}\,{\dot M}\, v_\mathrm{ff}^{-1}(r_\nu)\, B^2(r_\nu),
    \label{eq:L_radio_theory}
\end{equation}
where $\eta > 1$ is the cyclotron maser emission amplification factor, and $v_\mathrm{ff}$ and $B(r_\nu)$ are again taken at the emission radius $r_\nu$. The dimensionless coefficient $h = H/r$ still represents the radial width of the emission region. Obviously, $r_\nu$ is well below the Alfv{\'e}n radius, so from (\ref{eq:r_nu}) one gets $r_\nu \approx R(\nu_\mathrm{cyc}/\nu)^{1/3}$ or
\begin{equation}
    r_\nu \approx 11 R\, B_{\star,12}^{1/6} \nu_\mathrm{GHz}^{-1/6}.
\end{equation}
Substituting it into (\ref{eq:L_radio_theory}) one finally gets 
\begin{equation}
    L_\mathrm{ECME} \approx 9\times 10^{27}\,\eta h\,\dot M_{11}\, B_{\star,12}^{13/12}\,\nu_\mathrm{GHz}^{11/12}\, \mbox{ erg s}^{-1},
    \label{eq:lecme}
\end{equation}
so even for a moderately efficient maser amplification $\eta \gtrsim 1$ and $h \sim 1$ as assumed above, this estimation ends up with a similar luminosity as a purely mechanical quantity (\ref{eq:L_radio_max}). { Of course, both quantities linearly depend on $\dot M$. Thus, if the accreted mass rate is significantly lower than the Bondi rate, our estimates might be reduced appropriately.}

Note, however, the dependence of $L_\mathrm{ECME}$ on $\Bs$. 
For decaying magnetic fields, we expect that typical values are $\sim10^9$~G. Thus, either the luminosity is sometimes not sufficient to explain the observed sources, or large values of $\eta$ are necessary. 

%



\section{Conclusions}

We perform population synthesis calculations of isolated NSs in the Milky Way over $13.6$~Gyr using several models of the magnetic field behavior, the number density distribution of the ISM, and the spin evolution at the propeller stage. We assume that the turbulent moment can lead to the accretion disc formation around an isolated object and estimate the number of isolated accreting NSs with discs assuming the Kolmogorov scaling for the turbulent velocity in the ISM.

As a result, if the spin-down mechanism at the propeller stage is effective enough (models A, B, and C) so that a substantial fraction of the total galactic population of INSs can start accreting material from the ISM, there might be a few$\times10^5-10^6$ INSs with accretion discs.

In general, low characteristic velocities of the outer material relative to the NS and low magnetic field values favour disc formation. Thus, the dAINS population has velocities of $\lesssim20 \text{ km s}^{-1}$ if the field is constant, and $\lesssim40-50 \text{ km s}^{-1}$ if the field decays by four orders of magnitude over $13.6$~Gyr. The dAINS population consists of low-velocity neutron stars distributed near the Galactic plane, most of which have $|z|\lesssim30-240$~pc.

Adopting an MHD turbulence spectrum rather than a Kolmogorov one reduces the expected number of dAINSs with decaying magnetic fields by a factor of $10-100$, and eliminates dAINSs altogether if the magnetic field remains constant over $13.6$~Gyr. If the magnetic field decays, a few$\times10^3-10^5$ dAINSs can still be expected even in this less optimistic scenario.

We speculate that some of dAINSs can manifest themselves as long-period radio transient sources due to the emission mechanism similar to the one proposed by \cite{ferrario25}.

\section*{Acknowledgements}
The work of MDA and SBP (concept of the study, population synthesis of isolated neutron stars, and calculations of the fraction of accreting neutron stars with discs) was supported by the RSF grant
25-12-00012. 
AGS acknowledges funding from the European Union's Horizon MSCA-2022 research and innovation programme ``EinsteinWaves'' under grant agreement No. 101131233 and the Deutsche Forschungsgemeinschaft through individual research grant 570901071.  
We thank the anonymous referee for useful comments.




\bibliographystyle{elsarticle-harv} 
\bibliography{ains_disc} 







\end{document}